\documentclass[%
preprint,
bibnotes,
 amsmath,amssymb,
 aps,
prb,
floatfix,
]{revtex4-1}

\usepackage{graphicx,xcolor}
\usepackage{dcolumn}
\usepackage{bm}
\usepackage{gensymb}
\usepackage{hyphenat}
\usepackage{hyperref}

\usepackage{ulem}
\usepackage{etoolbox} 
\setcitestyle{square,open={},close={}}

\newcommand{\beginsupplement}{%
	\setcounter{table}{0}
	\renewcommand{\thetable}{S\arabic{table}}%
	\setcounter{figure}{0}
	\renewcommand{\thefigure}{S\arabic{figure}}%
}

\begin{document}


\title{Vibrational properties of sodosilicate glasses from first-principles calculations}

\author{Dimitrios Kilymis}
\altaffiliation{Current address: CIRIMAT, Universit\'e Toulouse 3 Paul Sabatier, CNRS, INPT, B\^at. CIRIMAT, 118, route de Narbonne, Toulouse cedex 9 31062, France.}
 \affiliation{%
 	Laboratoire Charles Coulomb (L2C), Univ. Montpellier, CNRS,
 	F-34095 Montpellier, France}%
\author{Simona Ispas}
\email{simona.ispas@umontpellier.fr}
\author{Bernard Hehlen}%
\email{bernard.hehlen@umontpellier.fr}
\affiliation{%
Laboratoire Charles Coulomb (L2C), Univ. Montpellier, CNRS,
Montpellier, France}%

\author{Sylvain Peuget}
\author{Jean-Marc Delaye}
\affiliation{CEA, DEN, DE2D SEVT LMPA F-30207 Bagnols-sur-C\`eze, France}%

\date{\today}

\begin{abstract}

The vibrational properties of three sodosilicate glasses have been investigated in the framework of density functional theory. 
The  pure vibrational density of states has been calculated for all systems and the different vibrational modes have been assigned to specific atoms or structural units. 
It is shown that the Na content
affects several vibrational features as the position and intensity of the $R$ band or the mixing of the rocking and bending atomic motions of the Si-O-Si bridges. 
The calculated Raman spectra have been found to agree with experimental observations and their decomposition indicated the dominant character of the non-bridging oxygen contribution on the spectra,
in particular for the high-frequency band, above 800~cm$^{-1}$. 
 The decomposition of the high-frequency Raman feature into vibrations of the depolymerized tetrahedra (i.e. $Q_n$-units) has revealed spectral shapes of the partial contributions that cannot be accounted for by simple Gaussians as frequently assumed 
in the treatment of experimentally obtained Raman spectra.

\end{abstract}


\maketitle

\section{\label{sec:intro}Introduction}

Raman and infra-red (IR) spectroscopies are well-established characterization tools of the silicate glasses and melts in both research laboratories and industries. 
Although the Raman and IR spectra are easy to collect, the spectral responses are complex,
combining broad and overlapping peaks due to the inherent structural disorder of these materials.
Consequently, the  decomposition and  extraction
of structural information from Raman and IR data are often based on parallelisms to well-known crystalline analogs, and result in a phenomenological and mostly qualitative approach~\cite{neuville2014advances,yadav2015review}. 
A good and quantitative understanding of
the structural significance of the features of the Raman spectra, as well as that of the two other  vibrational spectra
 (IR and inelastic neutron scattering spectroscopies) can be achieved
theoretically using atomistic calculations, in particular {\it ab initio} ones. 
In this context, the major aim of this study is to use atomistic simulation approaches in order to improve our
ability to connect measured vibrational properties of binary silicates to their structural
features.

The great advantage of the atomistic modeling  is that the positions of the atoms in the system under study are continuously monitored and the structure is exactly known. 
Therefore, correlations can be extracted between the atomic structure and the vibrational properties in order to arrive to a more quantitative interpretation of the vibrational properties and their dependence on the chemical composition, pressure, quench rates etc.
The latter is not directly accessible in experiments, but derived through functions describing how the modes with a given frequency are coupled to the probing radiation (photons, neutrons, etc.)~\cite{taraskin1997connection,Giacomazzi2009,thomas2013computing}.

The vibrational density of states (VDOS) can be in practice accessed either through the diagonalization of the dynamical matrix of the glass structure (in a local potential minimum), or the Fourier transform of the velocity autocorrelation function.  Within a classical approach,  most of the empirical potentials existing in the literature are quite reliable from the point of view of the structure  but they fail to reproduce the experimental VDOS,
even for systems as simple as pure silica or germania \cite{benoit2002,Huang2015}. \textit{Ab initio} methods are much more accurate in the prediction of the vibrational properties of materials but they have been scarcely used in this scope, due to the high computational cost associated with the treatment of glass systems~\cite{Sarnthein1997, Pasquarello1998, Pasquarello1998a,benoit2002, Giacomazzi2006, Giacomazzi2007, ganster2007structural, Tilocca2007, Pietrucci2008, Giacomazzi2009, Spiekermann2013, Pedesseau2015, Bhattarai2016}. More specifically, the fact that one has to model a system which lacks symmetry makes the calculations more time-consuming and finally results to systems that at best contain a few hundreds of atoms.

The calculations of the Raman spectra of a glass  
{have in general} three stages:  (i) generation of a structural model; (ii) the calculation of its vibrational properties; and (iii) the calculation of 
derivatives of the polarizability tensor with respect to atomic displacements (i.e., the Raman susceptibilities). For the latter, two frameworks can be considered, either a first-principles one~\cite{Lazzeri2003a,umari2005infrared} or an empirical one, called  bond polarizability model~\cite{wolkenstein1941intensities,alben1975vibrational}. 
Within the bond polarizability approximation, calculations for silica and  binary sodo-silicate glasses have achieved  satisfying  agreement with experimental data  \cite{zotov1999calculation,rahmani2003signature,Ispas2005,shcheblanov2015detailed}.  We note equally that Raman intensities can be equally  computed 
from the dynamical autocorrelation functions of the dielectric tensor in a  molecular dynamics (MD) simulation \cite{putrino2002anharmonic,thomas2013computing} approach which can be applied to liquids as well as to anharmonic solids, but is very demanding, requiring the calculation of the dielectric tensor
at each MD step.

Concerning the vibrational spectra of silicate glasses, most work carried out within 
an  \textit{ab initio} approach has focused on the simplest case of vitreous SiO$_2$. The early work of Sarnthein \textit{et al.} elucidated the nature of the high-frequency modes~\cite{Sarnthein1997}, while more recent works further elaborated on the VDOS of silica, as well as its IR and Raman spectra~\cite{Pasquarello1998, Pasquarello1998a,  Giacomazzi2009,  Bhattarai2016}. For more complex cases, theoretical calculations and analysis of the vibrational spectra have been carried out for sodosilicates~\cite{Ispas2005}, borosilicates\cite{Pedesseau2015}, aluminosilicates\cite{ganster2007structural, Bouyer2010}, phosphosilicates\cite{Tilocca2006, Tilocca2007, Corno2009} and magnesium-silicates\cite{Spiekermann2013}.

In this work, we report on the vibrational properties of three binary sodosilicate glasses whose structures have been prepared by combining classical and \textit{ab initio} MD simulations. We present their VDOS and its decompositions according to types of vibrations and contributions from specific structural units. At the same time we report results on the total and partial Raman spectra, aiming to help resolve the origins of different vibrational modes and better interpret experimental results.
As  Raman spectroscopy probes the short- and medium-range order in glasses, this modeling  approach may supply quantitative criteria for the spectral analysis. 


\section{\label{sec:simu_meth}Simulation Methods}

The selected sodo-silicate glasses of this study contain 20.0\%,  25.0\%, and  33.3\% Na$_2$O, hereafter called NS4, NS3 and NS2, respectively. Two samples per composition have been prepared, containing 180, 204, and 207 atoms for NS4, NS3, and NS2, respectively. In order to circumvent the costly preparation within 
an  \textit{ab initio}  molecular dynamics (MD) simulation, we have generated our structural models by employing classical MD approach during the melt-and-quench cycle followed by a rather short \textit{ab initio} MD run at room temperature. As it has been shown in past studies conducted using this combined approach~\cite{benoit2000,Ispas2001a,donadio2004photoelasticity,ganster2007structural,Giacomazzi2009,ispas2010structural}
the use of the \textit{ab initio} approach  mostly serves as a way to refine the short-range structure, while the medium range remains practically unchanged. The switch from a classical to an ab initio approach can have a dramatic effect on the calculated vibrational properties~\cite{benoit2002}. Similarly, Giacomazzi  \textit{et al.} extracted the vibrational properties of silica glass prepared with both methods with the results suggesting that they generally produce similar results, with some discrepancies arising due to the slightly different Si-O-Si angle distributions~\cite{Giacomazzi2009}.

 The initial preparation of the NS2 and NS3 glass structures, as well as  that of one  NS4 structure, has been performed using classical MD by employing the Guillot-Sator potential, with a cutoff set at 6~\AA \cite{Guillot2007}, starting from a random atomic distribution in a cubic box reproducing the experimental density~\cite{Bansal1986}. The systems were then heated to 4000~K and equilibrated for 20~ps using the constant volume-constant temperature (NVT) ensemble, before being quenched stepwise to 300~K with a rate of 10$^{11}$~K/s using NVT. Due to an 
 {inaccurate}
  description of the medium-range order in the case of the NS2, a third system has been prepared this time using a 10$^{10}$~K/s quench rate. Two final equilibrations of 20~ps and 5~ps were finally performed, using the constant pressure-constant temperature (NPT) ensemble and the constant volume-constant energy (NVE) ensemble, respectively. Due to slight contractions of the glasses during the NPT phase, some of the final structures were rescaled in order to match the experimental density values, and were then re-equilibrated for 5~ps using the NVE ensemble. The classical MD simulations were carried out with Berendsen thermostats and barostats, using an integration step equal to 1~fs. They were carried out using the DL\_POLY 4.01 software~\cite{Todorov2006}. 
As a second structure for the NS4 composition, we have used one previously generated, similarly through combined classical and {\it ab initio} MD simulations, albeit with a different classical potential~\cite{hehlen2017bimodal}.

In a second step, we used the above classical structural models as input configurations for {\it ab initio} MD simulations carried out using the Vienna {\it ab initio} package (\texttt{VASP})~\cite{vasp_code01,vasp_code02}.  The electronic degrees of freedom were treated in the framework of density functional theory (DFT), using the general gradient 
approximation (GGA) and  the PBEsol functional \cite{Perdew1996,PBEsol}. We started with an NVT run at room temperature for 2~ps, followed by an NVE run of same length. Complete computational details can be found in  previous publications~\cite{Pedesseau2015}. For the second NS4 structure, the \textit{ab initio} MD simulations have been performed using Car-Parrinello MD as implemented in \texttt{CPMD} code (see simulation details in Ref.~\cite[]{hehlen2017bimodal}). 

Further, in a third step, we have calculated the vibrational frequencies and eigenmodes, as well as the Raman tensors, using a DFT scheme as provided in the  \texttt{QUANTUM ESPRESSO} (QE) package (version 5.4)~\cite{Giannozzi2009}. The atomic positions have been initially relaxed using the BFGS minimizer with a fixed simulation box size setting the convergence criteria for the total energy and forces at $\mathrm{4\times10^{-8}}$~Ry and $\mathrm{10^{-5}}$~Ry/a.u., respectively. In case of a pressure higher than 1~GPa after the relaxation, the box was slightly suitably rescaled, keeping the final densities within 1.5\% of the experimental value.
The calculations were carried out using the GGA-PBE exchange-correlation functional~\cite{Perdew1996} with Troullier-Martins norm-conserving pseudopotentials~\cite{troullier1991efficient}. 
The performance of the pseudopotentials was assessed by investigating the structure and vibrational properties of two crystalline phyllosilicates, the sodium metasilicate (Na$_2$SiO$_3$) and natrosilite ($\alpha-$Na$_2$Si$_2$O$_5$). Calculations on the crystalline compounds were carried out using a kinetic energy cutoff of 100~Ry and Monkhorst-Pack $k-$point grids with a $\mathrm{4\times4\times4}$ and $\mathrm{3\times4\times3}$ resolution for the metasilicate and natrosilite, respectively. For the glasses, the same kinetic energy cutoff was used and calculations were carried out at the $\mathrm{\Gamma}$-point.
 {We should note at this point that, although two different exchange-correlation functionals have been used in the \textit{ab initio} MD runs, the final atomic relaxation has been carried out using PBE in all cases. Therefore, we consider that the final sets of positions used for the vibrational properties are consistent. As documented by 
	Ganster \textit{et al.}~\cite{ganster2007structural},  the \textit{ab initio}  runs at the later stages of the preparation of a glass model only serve as a way to refine the short-range order and do not affect the medium-range one. }

The vibrational properties (eigenfrequencies and eigenmodes)  were
calculated using the density-functional perturbation theory
method (DFPT) as implemented in QE code~\cite{Giannozzi2009,baroni2001phonons}.
All spectra for the NS3 and NS4 systems presented in the Sec.~\ref{sec:res} have been  averaged over the two models for the corresponding composition, after broadening the discrete spectra using Gaussians with a 20~cm$^{-1}$ FWHM, unless explicitly stated. In the case of NS2, the results only concern the system prepared with the slower quench rate (i.e.,~10$^{10}$~K/s).


\section{\label{sec:res}Results and Discussion}

\subsection{\label{sec:cryst}Crystalline compounds}

In Na$_2$SiO$_3$ and $\alpha-$Na$_2$Si$_2$O$_5$ the fourfold-fold coordinated Si atoms are connected to two and three bridging (i.e., double-coordinated) oxygens, respectively~\cite{Clark2001,Charpentier2004}, usually called  Q$_2$ and Q$_3$ units (we recall that  Q$_n$ species are tetrahedral environments with  $n=4,3,2,1,0$ BOs). Unit cells containing 12 and 36 atoms for Na$_2$SiO$_3$ and $\alpha-$Na$_2$Si$_2$O$_5$ were relaxed, with the results presented in Table~\ref{tab:tab1_cryst_struc}. The cell constants are predicted within 5\% compared to experiment \cite{McDonald1967,Pant1968a}. The intra-tetrahedral structural parameters are more accurately predicted than the inter-tetrahedral ones, whereas the largest deviations from the experimental values are observed in the case of Na-O bond lengths. The overestimation of these bond lengths is probably connected to the use of GGA/PBE approximation, as already reported in the literature~\cite{ishii2016dft}.

\begin{figure}[htbp]
\centering
\includegraphics[width=7cm]{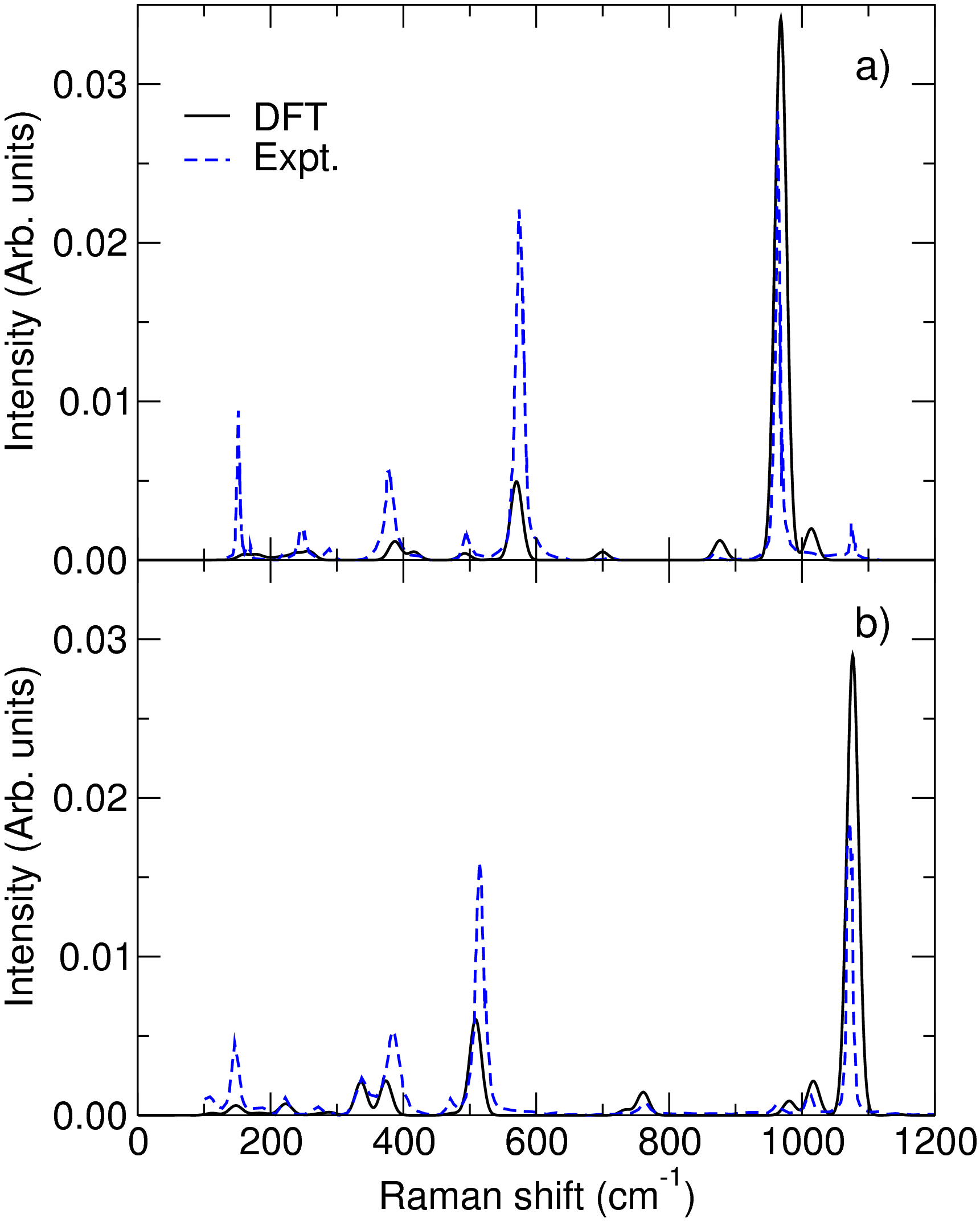}
\caption{\label{fig:raman_cryst} Polarized Raman spectra for a) Na$_2$SiO$_3$ and b) $\alpha-$Na$_2$Si$_2$O$_5$. Experimental spectra are from Refs.~\cite[]{Richet1996,You2001}.}
\end{figure}

{\small 
\begin{table} [htbp]
\begin{ruledtabular}
\caption{\label{tab:tab1_cryst_struc} Cell size, average values of the bond lengths (Si-BO, Si-NBO and Na-O) and angles (Si-O-Si and O-Si-O) formed by the unit cell atoms for Na$_2$SiO$_3$ and $\alpha-$Na$_2$Si$_2$O$_5$. Average percent errors for the angles and bond lengths are given in parentheses.}
\begin{center}
\begin{tabular}{lcccc}
& Na$_2$SiO$_3$ & Exp.$^a$ & $\alpha-$Na$_2$Si$_2$O$_5$ & Exp.$^b$\\
\hline
$a$ (\AA) & 10.92 & 10.48 & 6.65 & 6.409\\
$b$ (\AA) & 6.34 & 6.07 & 15.83 & 15.422\\
$c$ (\AA) & 4.97 & 4.82 & 5.02 & 4.896\\
\hline 
Si-O-Si~($^\circ$) & 139.2 & 133.7 & 162.1 & 160.0\\
                 &       &       & 144.2 & 138.9\\
 		& (4.1)	& 		& (2.6) &\\
 O-Si-O~($^\circ$) & 103.9 & 103.1 & 108.6 & 109.1\\
                & 110.7 & 110.8 & 107.5 & 107.5\\
                & 116.5 & 116.9 & 112.9 & 113.2\\
                & 107.0 & 107.1 & 106.0 & 105.4\\
                &       &       & 108.6 & 108.2\\
                &       &       & 113.0 & 113.0\\
			& (0.1)	&		& (0.0) &\\
\hline
Si-BO~(\AA) & 1.689 & 1.677 & 1.621 & 1.609\\
         & 1.675 & 1.668 & 1.651 & 1.643\\
         &		&		& 1.646 & 1.638\\
 & (0.6)	&		& (0.6)	&\\
Si-NBO~(\AA) & 1.606 & 1.592 & 1.590 & 1.578\\
 		& (0.9)	&		& (0.8)	&\\
Na-O~(\AA) & 2.436 & 2.282 & 2.536 & 2.386\\
        & 2.450 & 2.303 & 2.468 & 2.338\\
        & 2.609 & 2.549 & 2.480 & 2.373\\
        & 2.483 & 2.370 & 2.444 & 2.290\\
        & 2.550 & 2.404 & 2.629 & 2.600\\
 & (5.3)	&		& (4.8)	& \\
\hline
\multicolumn{2}{l}{$^a$ Ref.~\cite[]{McDonald1967}; $^b$ Ref.~\cite[]{Pant1968a} }
\end{tabular}

\end{center}
\end{ruledtabular}
\end{table}

}
The Raman spectra for the two systems are given in Fig.~\ref{fig:raman_cryst}. 
{As there is a constant underestimation of the frequencies of the calculated spectra, the presented values have been multiplied by a factor of 1.04.}
For both crystal structures, we observe a good agreement of the band positions with respect to available experimental data~\cite{Brawer1975,Richet1996,You2001,Kato2004}, except from the weak Si-O stretching band of Na$_2$SiO$_3$ at 1063~cm$^{-1}$ which is not present in our calculations. However, the overall good description of the spectra suggests that the chosen pseudopotentials and simulation parameters are adequate for the calculation of the spectra for the glasses.

\subsection{\label{sec:nsx_struct}Structure of the NSX glasses}

\begin{figure} [htbp]
\centering
\includegraphics[width=10cm]{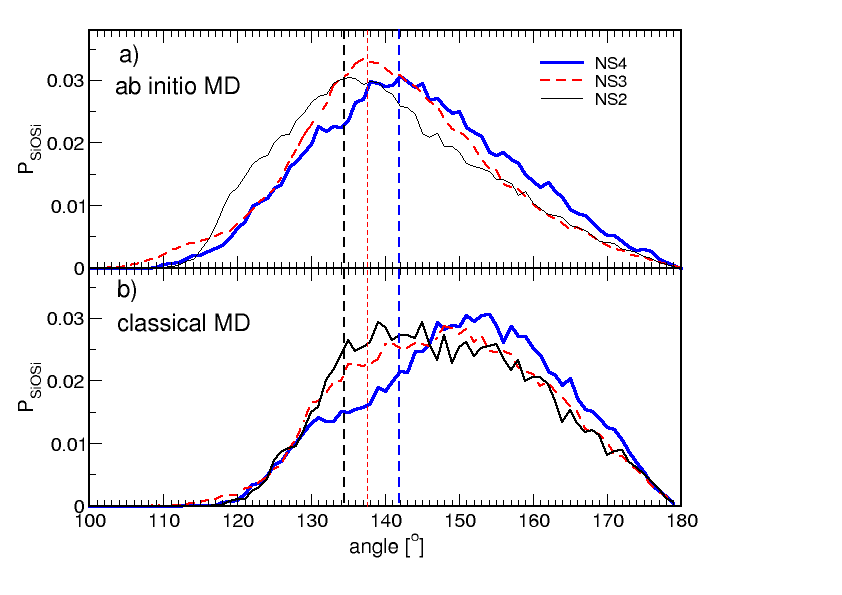} 
\caption{\label{fig:siosi_compMD} SiOSi BADs of the simulated NS4, NS3, and NS2 glassy samples at 300~K: (a) (Top) Extracted from \texttt{VASP} ab initio MD runs; (b) (Bottom)  Extracted from \texttt{DL$_-$POLY} classical MD runs.
	}
\end{figure}

\begin{figure} [htbp]
\centering
\includegraphics[width=8cm]{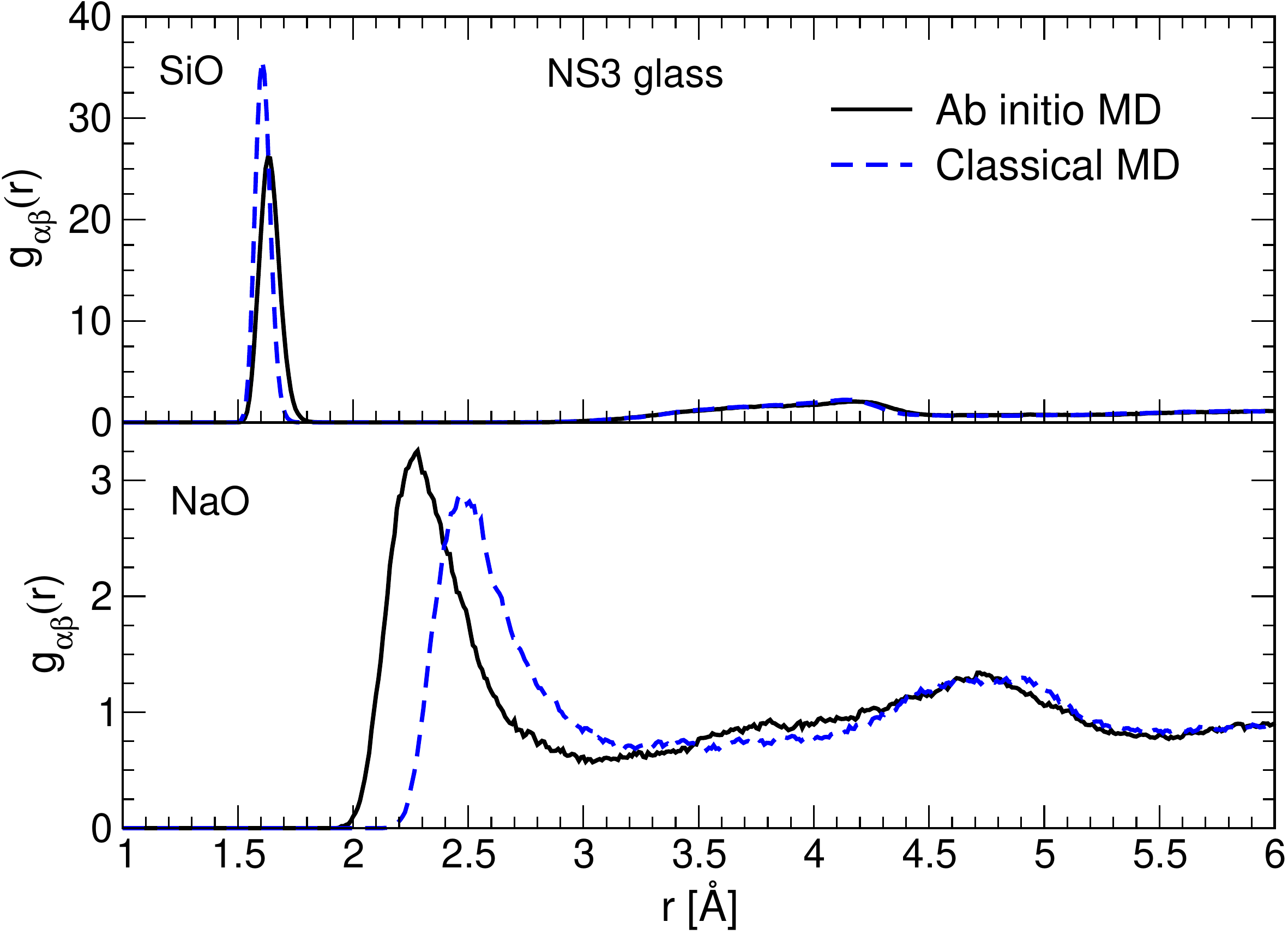}
\caption{\label{fig:sionao_comp} Radial distribution functions for SiO and NaO pairs computed for the classical and \textit{ab initio} runs, at 300¬K and for NS3 glass structures.}
\end{figure}

When switching from classical to {\it ab initio} MD, we have observed changes of some structural features such as the shift 
to smaller angles shown by the SiOSi bond angle distribution (BAD) (see Fig.~\ref{fig:siosi_compMD}) or the shifts of the first peaks of SiO and NaO radial distribution functions (RDF) plotted in Fig.~\ref{fig:sionao_comp} for NS3. For  the SiOSi BADs plotted in Fig.~\ref{fig:siosi_compMD}, their average value does not strongly depend on the sodium content (see Table~\ref{tab:tab2_glass_struct}), but their maximum does, as its location shifts to lower values   with increasing Na content. 
{This trend is consistent with experimental observations and is particularly apparent for the {\it ab initio} data~\cite{hehlen2017bimodal}.}
For the SiO RDF, the first peak  shifts to larger values together with a broadening, which may reflect the deformation of SiO$_4$ tetrahedra bearing NBOs, known to have shorter bond lengths with Si atoms compared to Si-BO bonds~\cite{Ispas2001a, Pedesseau2015}. Concerning the Na-O RDF, Fig.~\ref{fig:sionao_comp} shows a decreasing of NaO bond lengths, with an average NaO bond length moving from  2.73~\AA\, during the classical simulation to 2.50~\AA\, for the {\it ab initio} trajectory. This variation improves the comparison with experimental values (see  Table~\ref{tab:tab2_glass_struct}), but overall it seems that, as for the crystal structures,  these bond lengths are overestimated, reflecting the already mentioned  trend of {\it ab initio} calculations using the GGA approximation.

In Table~\ref{tab:tab2_glass_struct} we summarize some structural parameters of the three glass models used for the vibrational calculations (i.e., relaxed to $0$~K). The amount of NBO in the structures was found to be in very close agreement to available NMR results~\cite{Maekawa1991,Angeli2011,Lee2003}. The percentages of the 
NBOs are accurately reproduced in all glasses, while  the   agreement of the  Q$_n$ species with experimental results is better for the well-polymerized NS4 glass than for NS2. Overall, one can state that the short-range structural features of our models compare reasonably well with experiments, but deviations exist in terms of distribution of the Q$_n$ species, and thus for the medium-range structure.
 This discrepancy can be attributed to the choice of the empirical potential during the initial preparation of the glass models, since the \textit{ab initio} equilibration at room temperature and the subsequent energy minimization produced negligible changes in the medium-range structure. It can be also linked to the employed high quench rate from the liquid. 

\begin{table*} [htbp]
\caption{\label{tab:tab2_glass_struct} Density, NBO percentage, and Q$_n$ species percentages for the NSX glasses, as well average values of some bond lengths and SiOSi angle for the glass models used for vibrational calculations.}
\begin{ruledtabular}

\begin{tabular}{l|cc|cc|cc}

& \multicolumn{2}{c|}{NS4} &\multicolumn{2}{c|}{NS3}&\multicolumn{2}{c}{NS2}\\
& Sim. & Exp. & Sim. & Exp. & Sim. & Exp.\\
\hline
$\rho$ (g/cm$^3$) & 2.38 & 2.38${}^a$ & 2.40 & 2.43${}^a$ & 2.48 & 2.49${}^a$\\
\% NBO & 21.8 & 23.1${}^b$ & 28.6 & 27.0${}^b$, 28.0$^c$ ,28.0$^d$   & 40.0 & 40.0${}^b$\\
\% Q$_1$ &  &  &  &  & 2.2 & - \\
\% Q$_2$ & 8.3 & 3.0${}^b$ , 0.8$^e$ & 11.8 & 1.0${}^b$ , 2.3$^c$   2.4$^e$ , 1.4$^f$  & 17.4 & 10.0${}^b$ , 8.4$^f$  \\
\% Q$_3$ & 32.3 & 48.0${}^b$ , 50.0$^e$ & 43.1 & 61.0${}^b$ , 60.5$^c$  , 62.3$^e$ , 34.9$^f$ & 58.7 & 79.0$^b$ , 81.0$^f$ \\
\% Q$_4$ & 59.4 & 50.0$^b$  49.2$^e$ & 45.1 & 38.0$^b$ 37.2$^c$  35.3$^e$ & 21.7  & 11.0$^b$ 10.3$^f$  \\
\hline
Si-O~(\AA) &  1.627 & 1.617$^g$ &		1.628	&1.62$^g$	&			1.632	&		 1.61-1.62${}^h$, 1.631${}^i$\\
Si-BO~(\AA)&  1.635	& &			1.638		& &							1.649\\
Si-NBO~(\AA)& 1.574		& &		1.581		& &						1.587 \\
Na-O~(\AA)&  2.522		& 2.30${}^h$ &		2.512			&  	&	2.551  & 2.30-2.36${}^h$\\
SiOSi~($^\circ$) & 143.08		& &		142.45		& &						143.49
\end{tabular}

${}^a$~Ref.\cite[]{Bansal1986};
${}^b$~Ref.\cite[]{Maekawa1991};
${}^c$ Ref.\cite[]{Angeli2011};
${}^d$ Ref.\cite[]{Lee2003};
${}^e$ Ref.\cite[]{Olivier2001}; ${}^f$ Ref.\cite[]{Malfait2008}
${}^g$ Ref.\cite[]{henderson1995si}; ${}^h$ Ref.\cite[]{greaves1981local}, ${}^i$ Ref.\cite[]{misawa1980short}

\end{ruledtabular}

\end{table*}

It is also interesting to analyze the intertetrahedral Si-O-Si angles with respect to the  Q$_n$ speciation of the two silicons.  In this scope, we have averaged the corresponding values over all five glass models, wherever sufficient statistics were available, and found the largest angles being formed by Q$_4$-O-Q$_4$ (144.2$^\circ$), followed by Q$_4$-O-Q$_3$ (143.8$^\circ$), Q$_4$-O-Q$_2$ (141.5$^\circ$), Q$_3$-O-Q$_3$ (140.1$^\circ$), and finally Q$_3$-O-Q$_2$ (137.3$^\circ$). This change in the Q$_m$-O-Q$_{n}$ angle could be well related with a steric effect caused by the increased amount of Na around the Q$_n$ entities when increasing the number of NBOs. 
We have hence calculated  
the average number of Na atoms around a Q${}_n$ unit (i.e., the Na atoms around the four oxygens and within a sphere of radius equal to the first O-Na coordination shell), and we have found increasing values when going from Q$_4$ to Q$_2$, for a given composition, and when increasing the soda content.    For example, for Q${}_3$ species, we have found  $\approx 3.8$ sodiums for NS4, $\approx 4.4$ sodiums for NS3 and finally $\approx 5.8$ for NS2 composition. Further, by calculating  the average number of Na around each NBO, we have found values equal to 2.9, 3.2, and 3.4 for NS4, NS3, and NS2, respectively, similarly to previous \textit{ab initio} calculations on NS4 and a sodium borosilicate glass~\cite{Charpentier2004,Pedesseau2015}. 
Of course, this implies that more than one alkali atom is found in the vicinity of each NBO, therefore deviating from the traditional theoretical model for the structure of sodosilicates, which assumes a one-to-one correspondence.


\subsection{\label{sec:vdos}Vibrational density of states}

As stated in Sec.~\ref{sec:simu_meth}, the dynamical matrix of each glass model was computed within the framework of DFPT, and 
then its diagonalization yielded the $3N$ eigenfrequencies  $\omega_p$ and their corresponding normalized $3N$-component eigenvectors $\mathbf{e}(\omega_p)$  ($p=1,2,3,\ldots ,~ 3N$), where $N$ is the number of atoms in the system.
The total VDOS has been then computed as:
\begin{equation}
g\left(\omega\right)=\frac{1}{3N-3}\sum_{p=4}^{3N}\delta\left(\omega-\omega_p\right)
\end{equation}
The knowledge of the eigenfrequencies and eigenvectors makes possible further calculations, starting with the so-called partial VDOS corresponding to the contributions of different groups of atoms (as, for example, the species):
\begin{equation}
g_\alpha\left(\omega\right)=\frac{1}{3N-3}\sum_{p=4}^{3N}
\sum_{I=1}^{N_\alpha}\sum_{k=1}^3{|\textbf{e}_{I,k}\left(\omega_p\right)|}^2\delta\left(\omega-\omega_p\right)
\label{eq:pvdos}
\end{equation}
where $N_\alpha$ is the number of atoms in the group labeled $\alpha$  and $\textbf{e}_{I,k}\left(\omega_p\right)$ are the $3$-components  of the eigenvector $\mathbf{e}(\omega_p)$  with the displacement of the particle $I$. 

\begin{figure} [htbp]
\centering
\includegraphics[width=14cm]{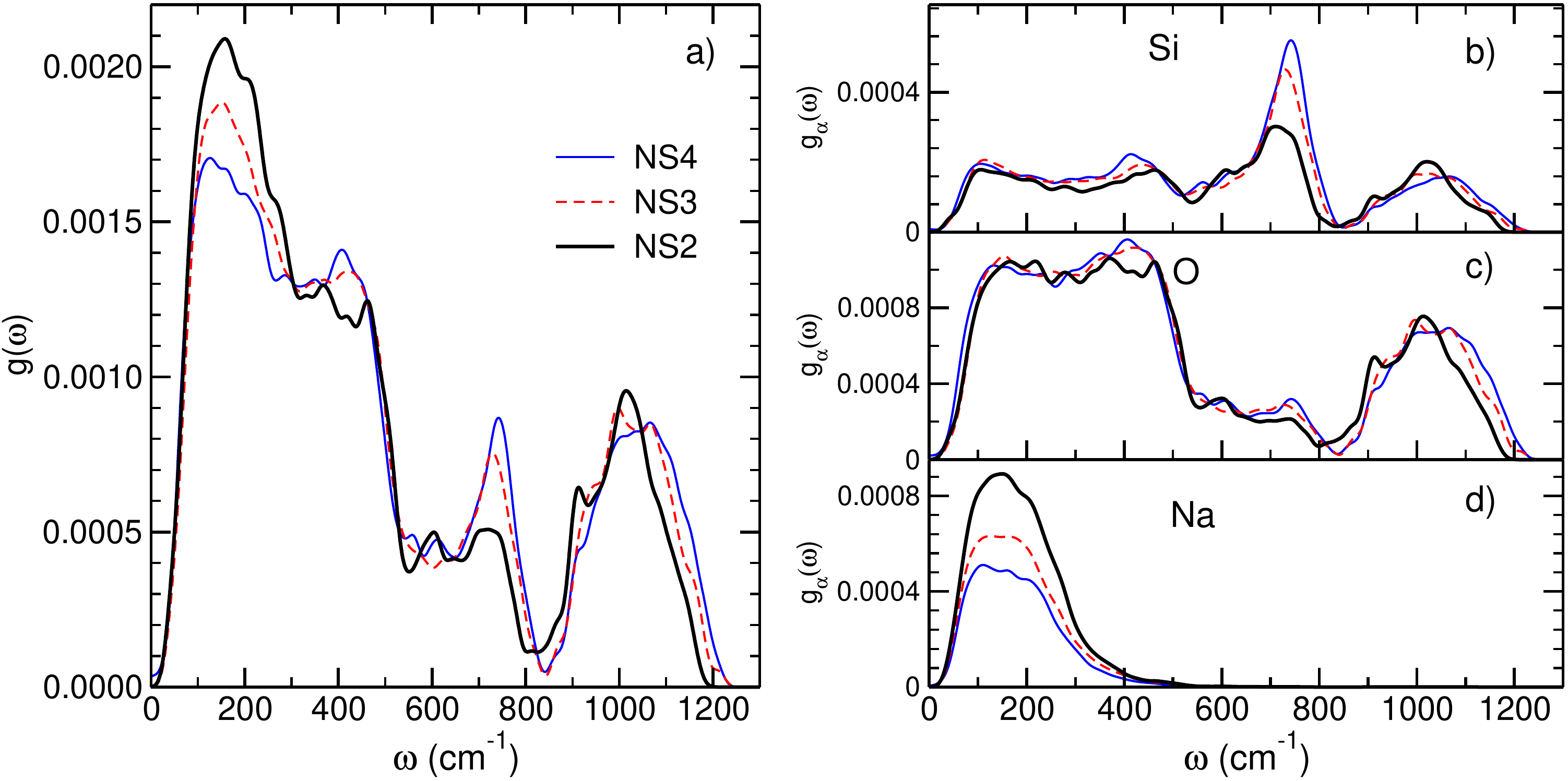}
\caption{\label{fig:fig2-vdos_all} (a) Total VDOSs of the NSX glasses and partial contributions from (b) Si,(c) O and (d) Na atoms. The total VDOSs were normalized to unity, while the partial VDOS for a given composition were normalized so that their sum  gives the corresponding total VDOS. }
\end{figure}

The total VDOSs  of NS4, NS3 and NS2 glasses are shown in Fig.~\ref{fig:fig2-vdos_all} alongside the  partial contributions of the three species $\alpha=$~Si, O, Na. Firstly we can identify three main bands with  features (position, shape and intensity) changing when the Na content varies. The presence of three bands as well as qualitatively similar  decompositions with respect to the contributions from the three species have been already reported in previous theoretical studies for NS4 glass  using  classical and \textit{ab initio}~\cite{zotov1999calculation,Ispas2005}, as well as by  fully \textit{ab initio} studies of more complex silicate glasses~\cite{tilocca2006ab,Pedesseau2015}.
For the low frequency range below 300~cm$^{-1}$, we observe an intensity increase  when going from NS4 to NS2, 
which is directly correlated to the intensity increase presented by the Na partial VDOSs when the Na sodium content increases [see Fig.~\ref{fig:fig2-vdos_all}(d)]. We note that the Na contributions are situated only in the bottom part of the spectra below 400~cm$^{-1}$, as already pointed out in previous calculations for silicates 
containing sodium~\cite{Ispas2005,tilocca2006ab,Pedesseau2015}. In terms of relative intensity of the three partials, it is evident that the O contribution dominates the VDOS [see Figs.~\ref{fig:fig2-vdos_all}(b)--\ref{fig:fig2-vdos_all}(d)].  
 To our knowledge, the only available experimental data in order to compare the VDOS come from  heat-capacity calculations  and concern the lower end of the spectra~\cite{richet2009heat}, and show an intensity increase with increasing sodium content.

\begin{figure} [htbp]
\centering
\includegraphics[width=10cm]{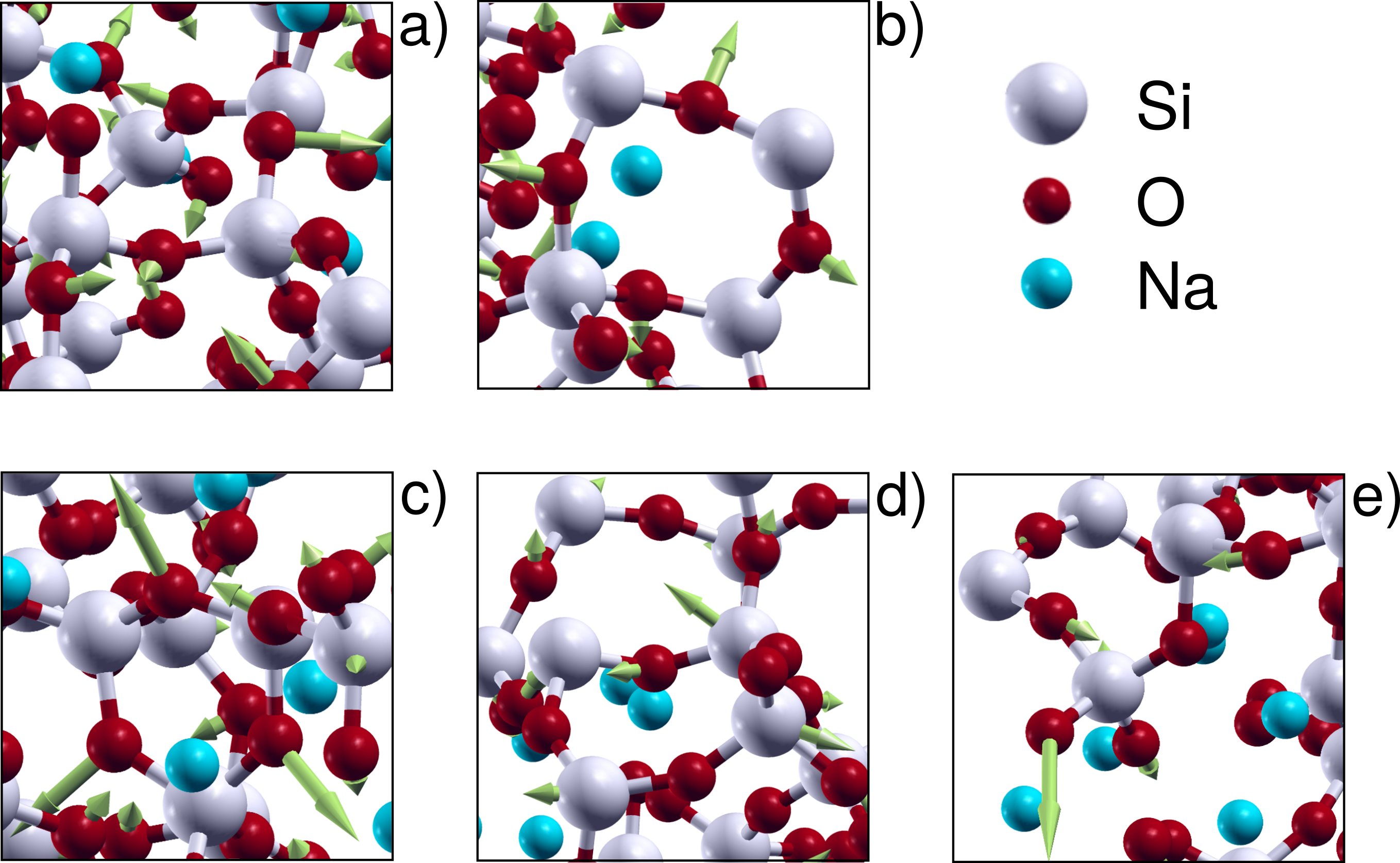} 
\caption{\label{fig:snaps} Visualization of the vibrational modes at (a) 416~cm$^{-1}$, (b)  481~cm$^{-1}$, (c)  602~cm$^{-1}$, (d) 694~cm$^{-1}$ and (e) 976~cm$^{-1}$ for the NS4 glasses. White atoms are Si, red are O, and blue are Na. For the color version, the reader is referred to the web version of the paper.}
\end{figure}

The change in the Na amount also gives rise to modifications of the total VDOSs for the frequency ranges above $400$~cm$^{-1}$. 
We notice an intensity decrease with increasing Na concentration in the range $ 400-500$~cm$^{-1}$ [see Fig.~\ref{fig:fig2-vdos_all}(a)],  a feature that originates from the changes presented by both  oxygen and silicon partial VDOSs [see Figs.~\ref{fig:fig2-vdos_all}(b) and ~\ref{fig:fig2-vdos_all}(c)] and can be correlated to changes in the glass polymerization due to the modifier atoms. 
Further on, there is a peak pointing at $\approx 740$~cm$^{-1}$ which is strongly affected by the Na content:  its  intensity decreases with increasing Na amount, alongside a slight shift to lower frequencies.  Again, we can link these variations to changes shown by both Si and O partial VDOS, and especially the ones of  the silicons with a decrease of the maximum intensity of 40\% between NS4 and NS2 samples, as shown in Fig.~\ref{fig:fig2-vdos_all}(b).  Finally,  the three VDOSs are characterized by a broad band  approximatively above 850~cm$^{-1}$ and up to 1250~cm$^{-1}$,  with contributions from Si and O atoms which shifts to lower frequencies with increasing Na content [see Figs.~\ref{fig:fig2-vdos_all}(a)--\ref{fig:fig2-vdos_all}(c)]. This band  is of course reminiscent of the so-called high-frequency band  in pure silica, and its softening due to the silicate network depolymerization has been previously  reported in \textit{ab initio} studies for other silicates~\cite{tilocca2006ab,Spiekermann2013,Pedesseau2015}.   The assignment of this band and the relation with the network connectivity  will be discussed in details in the next subsections.

We can initially proceed to a  visual inspection of the vibrational modes (videos are provided as Supplemental Material~\cite{suppl_mat}).  The nature of the vibrations was found to be qualitatively similar in all three glasses under study. 
Figure~\ref{fig:snaps} depicts some characteristic snapshots taken from the NS4 models and visualized using XCrySDen~\cite{Kokalj2003}.
For  the low-end of the VDOS we observe mainly motions of rigid tetrahedra together with neighboring sodium atoms.  Further, for increasing frequencies up to 350~cm$^{-1}$, we first see   dangling and then stretching motions of the NBO-Na bonds.  
At the same time, the Si-O-Si bridges exhibit a combined rocking/bending motion  [Fig.~\ref{fig:snaps}(a)].
At higher frequencies we can also observe characteristic collective motions like the breathing modes of 4-member and 3-member silicate rings,
shown in Figs.~\ref{fig:snaps}(b) and \ref{fig:snaps}(c). 
In the region 700-800~cm$^{-1}$ [see Fig.\ref{fig:snaps}(d)], we still have the mixed rocking/bending motions of Si-O-Si bridges but we equally notice  an increased contribution stemming from Si atoms, which corroborates the above discussed features of the Si partial VDOS. At a lower extent, we can still identify the presence of a mixed motion on NBOs combining a stretching of the Si-NBO bond together with a motion orthogonal to the  Si-NBO direction. Finally, in the high frequency band above 850~cm$^{-1}$  [see Fig.\ref{fig:snaps}(e)], we mostly observe the  stretchings of both Si-BO and Si-NBO bonds, as  previously reported for a NS4 glass~\cite{zotov1999calculation} within a classical approach  or for magnesio-silicate glasses~\cite{Spiekermann2013} within a DFT calculation.

\begin{figure}[htbp]
\centering  
\includegraphics[width=12cm]{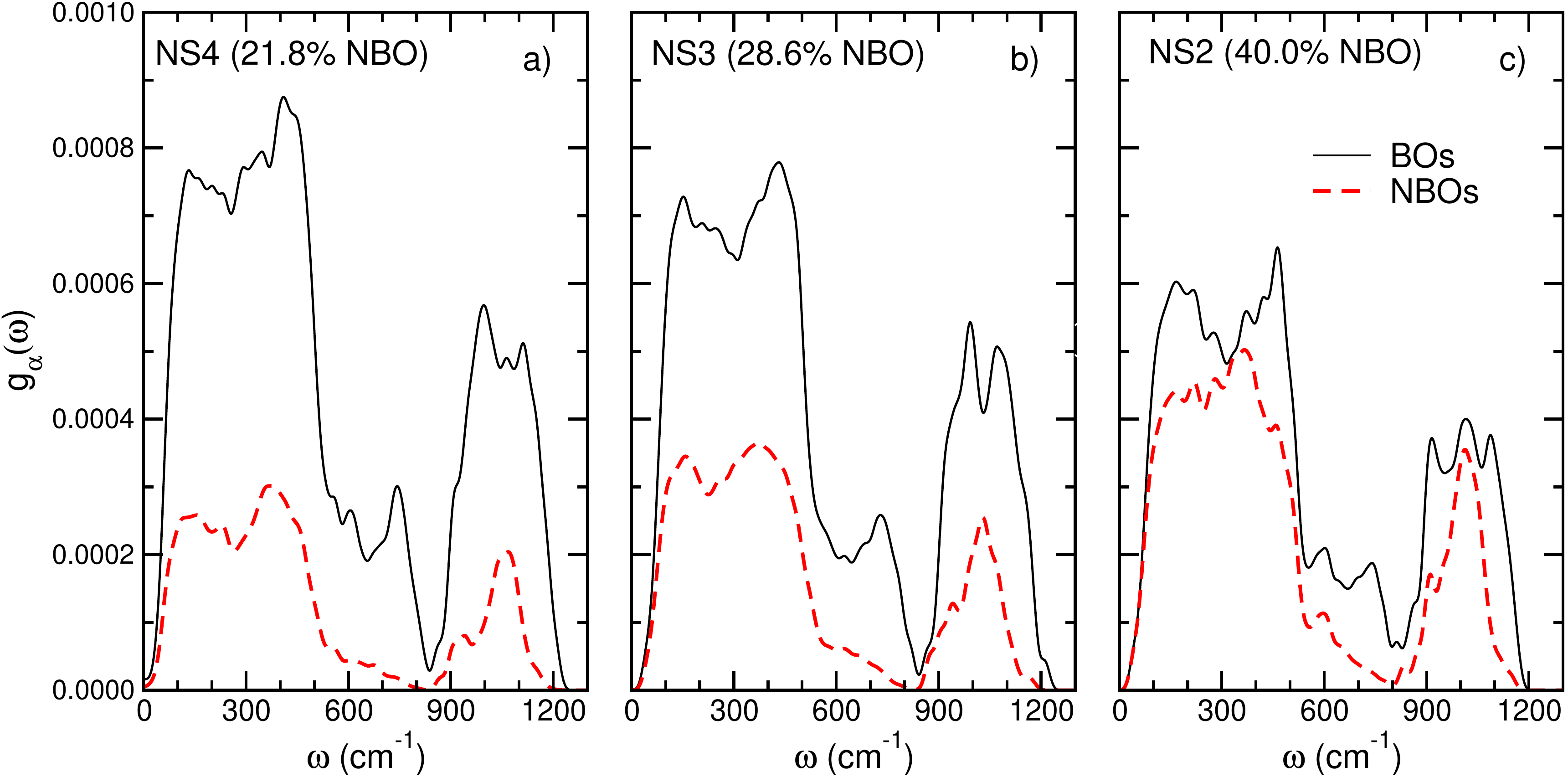}
\caption{\label{fig:vdos_bo-nbo} Contribution to the VDOS of BO (solid lines) and NBO (dashed lines) for (a) NS4, (b) NS3, and (c) NS2.}
\end{figure}

After obtaining a qualitative overview of the vibrational modes of the NSX glasses, we can proceed to a more quantitative  analysis.
As mentioned earlier, the most intense contribution to the VDOS stems from the O atoms. In Fig.~\ref{fig:vdos_bo-nbo}, we present the decomposition of the O partial VDOS with respect to BOs and NBOs contributions. The results naturally show the relative decrease of the BO contribution in the entire spectrum, as the glasses become more depolymerized, and the simultaneous increase of the NBO contribution. We notice that around 700~cm$^{-1}$ the BO contribution  remains dominant, even in the case of the NS2 glass,  and, as we will see below,  the  vibrational modes in this region of the spectra result from a superposition of  rockings/bendings of the Si-O-Si bridgings.

\begin{figure} [htbp]
\centering 
\includegraphics[width=8cm]{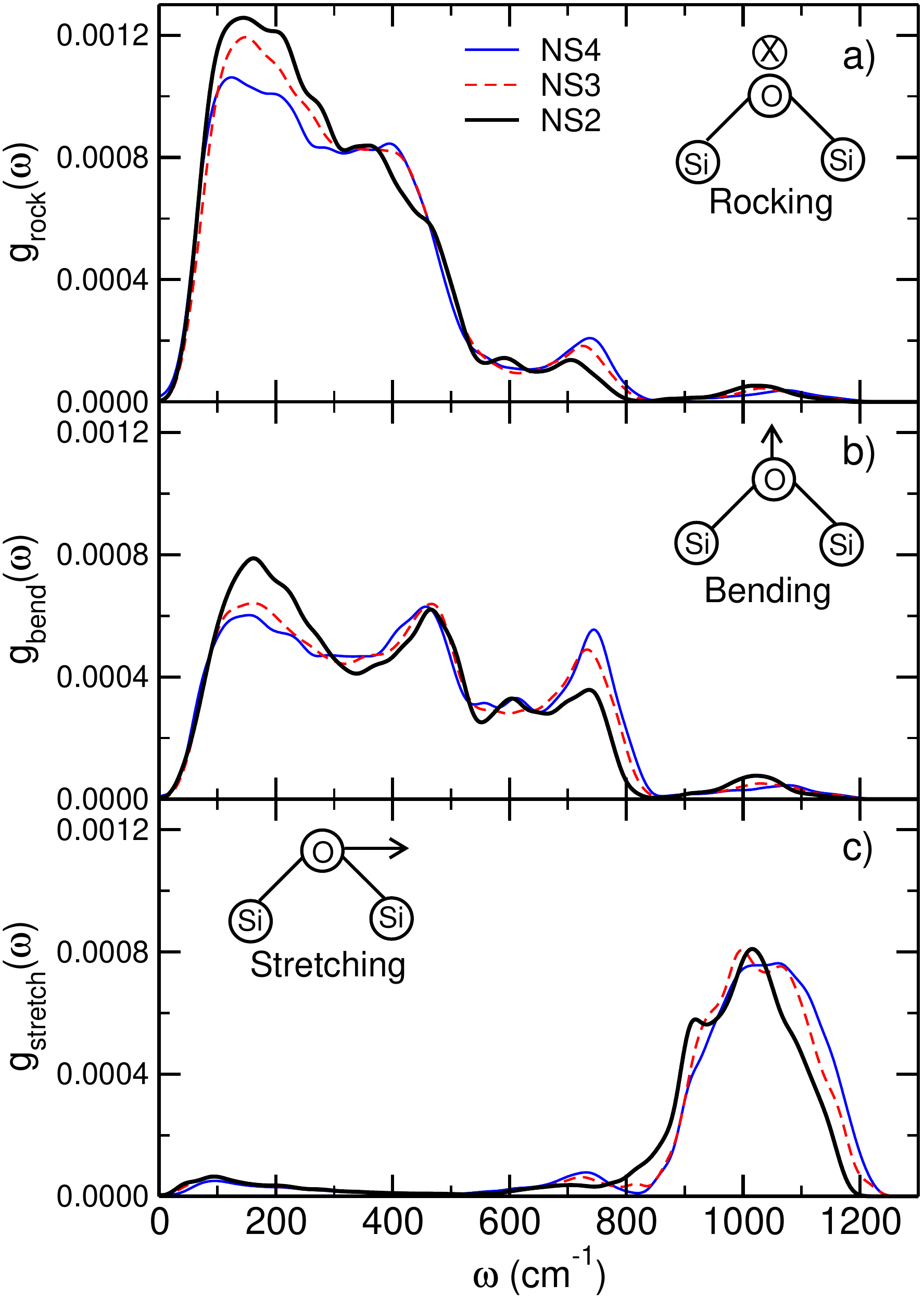}
\includegraphics[width=7.5cm]{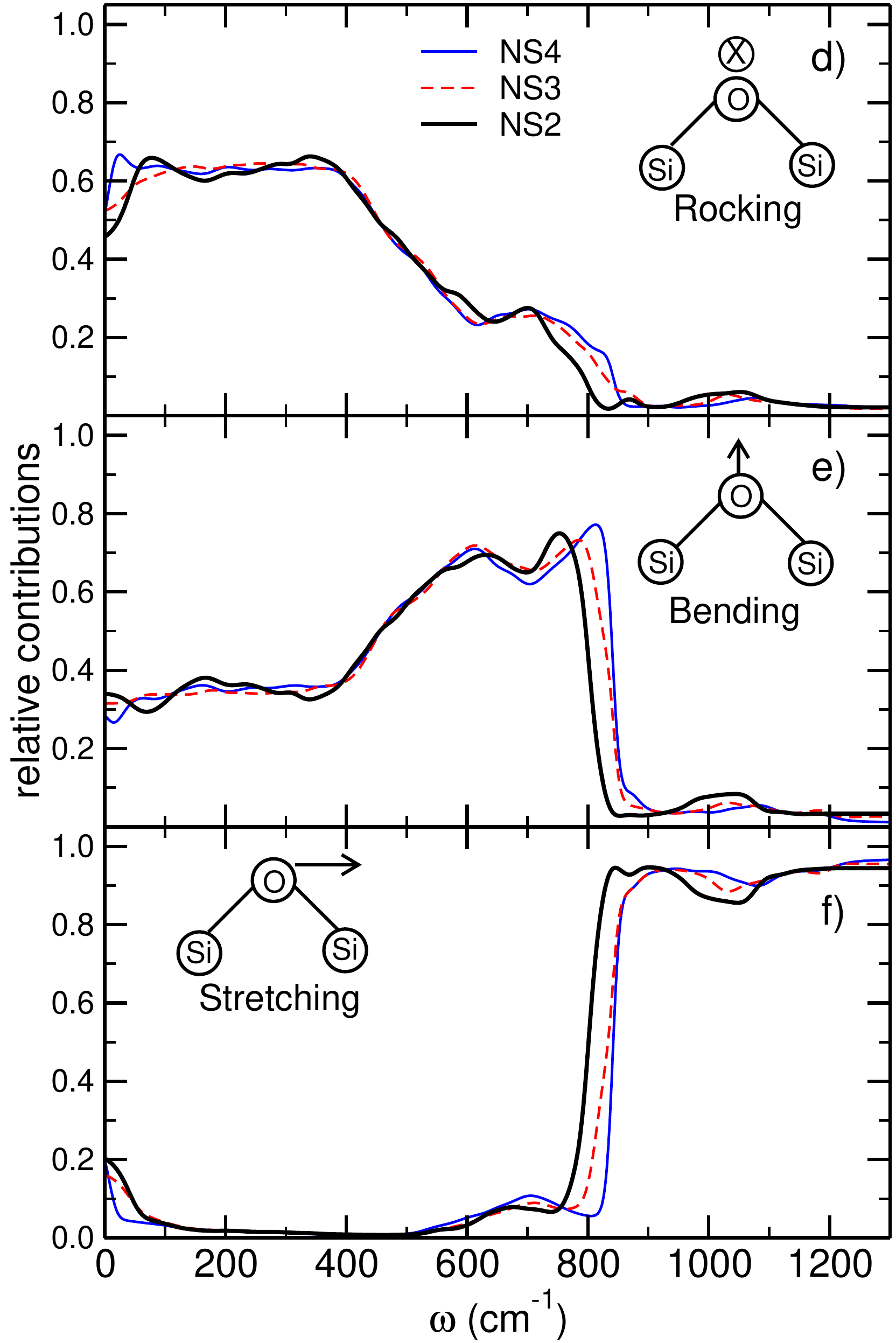}
\caption{\label{fig:vdos_rbs} Frequency and composition dependencies of the (a) rocking, (b) bending, and (c) stretching components of BO modes in the three NSX glasses (X=4,3,2). (d)--(f) show the relative contributions of the rocking, bending and stretching motions, respectively, with respect to the total VDOS. }
\end{figure}

We have further analysed the vibrational modes of the Si-O-Si bridges according to the decomposition of the BO motions into rocking, bending, and stretching components. 
For each mode, we have calculated the relative displacement of the BO with respect to its two neighboring Si atoms and 
projected it over three orthogonal directions~\cite{Taraskin1997}: the rocking direction orthogonal to the Si–O–Si plane, the bending direction along the bisector of the Si-O-Si angle and finally the stretching direction orthogonal to the two previous ones.  The corresponding partial VDOS are presented in 
Figs.~\ref{fig:vdos_rbs}(a)--\ref{fig:vdos_rbs}(c), while Figs.~\ref{fig:vdos_rbs}(d)--\ref{fig:vdos_rbs}(f) shows the relative contributions of rocking, bending, and stretching motions, respectively, with respect to  the  total VDOS. The data plotted in Fig.~\ref{fig:vdos_rbs}  show a mix of rocking and bending motions in the low- and mid-frequency range, with the former being more intense in the region between 100-400~cm$^{-1}$, whereas the latter are relatively more pronounced between 500-800~cm$^{-1}$. On the contrary, the high frequency band contains essentially stretching modes of BO. 
Regarding the effect of the glass composition, the increase of Na amount when going from NS4 to NS2  leads to an increase of the 
contributions from rocking and bending modes at the lower part of the VDOS (below 700~cm$^{-1}$). At the same time, their intensities for the band around 750~cm$^{-1}$ decrease, and also shift to lower frequencies (the effect being more pronounced for the relative contributions, as shown in Figs.~\ref{fig:vdos_rbs}(d)--\ref{fig:vdos_rbs}(e). The increase of Na has also a limited effect 
on the intensity of the high-frequency band which only shifts to lower frequencies 
[see Fig.~\ref{fig:vdos_rbs}(e)]. 
Similar decompositions of the oxygen motions have been previously carried out  for pure silica described using an effective pair potential~\cite{Taraskin1997,shcheblanov2015detailed} or within an \textit{ab initio} framework~\cite{Pasquarello1998a,umari2003first}, as well as for the NS4 glass using a harmonic potential~\cite{zotov1999calculation}. These studies have revealed the same strong dependence of BO motions with respect to the frequency, and a qualitative agreement  with respect to the relative ratios between the three contributions.

\begin{figure} [htbp]
\centering  
\includegraphics[width=7cm]{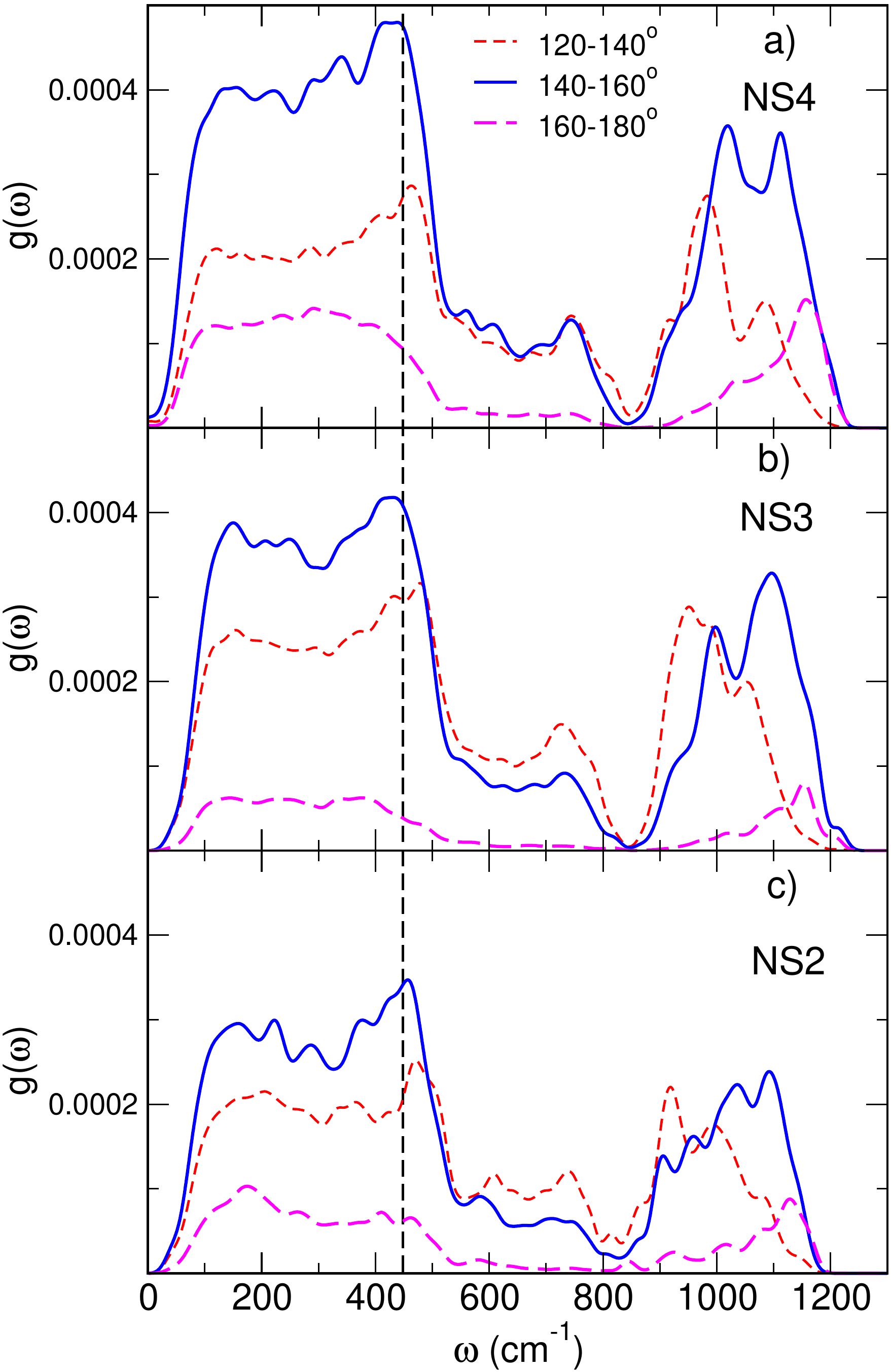}  
\caption{\label{fig:vdos_ang} Decomposition of the BOs  contribution with respect to  their Si-BO-Si angle  for (a) NS4, (b) NS3, and (c) NS2.}
\end{figure}

\begin{figure} [htbp]
\centering
\includegraphics[width=7cm]{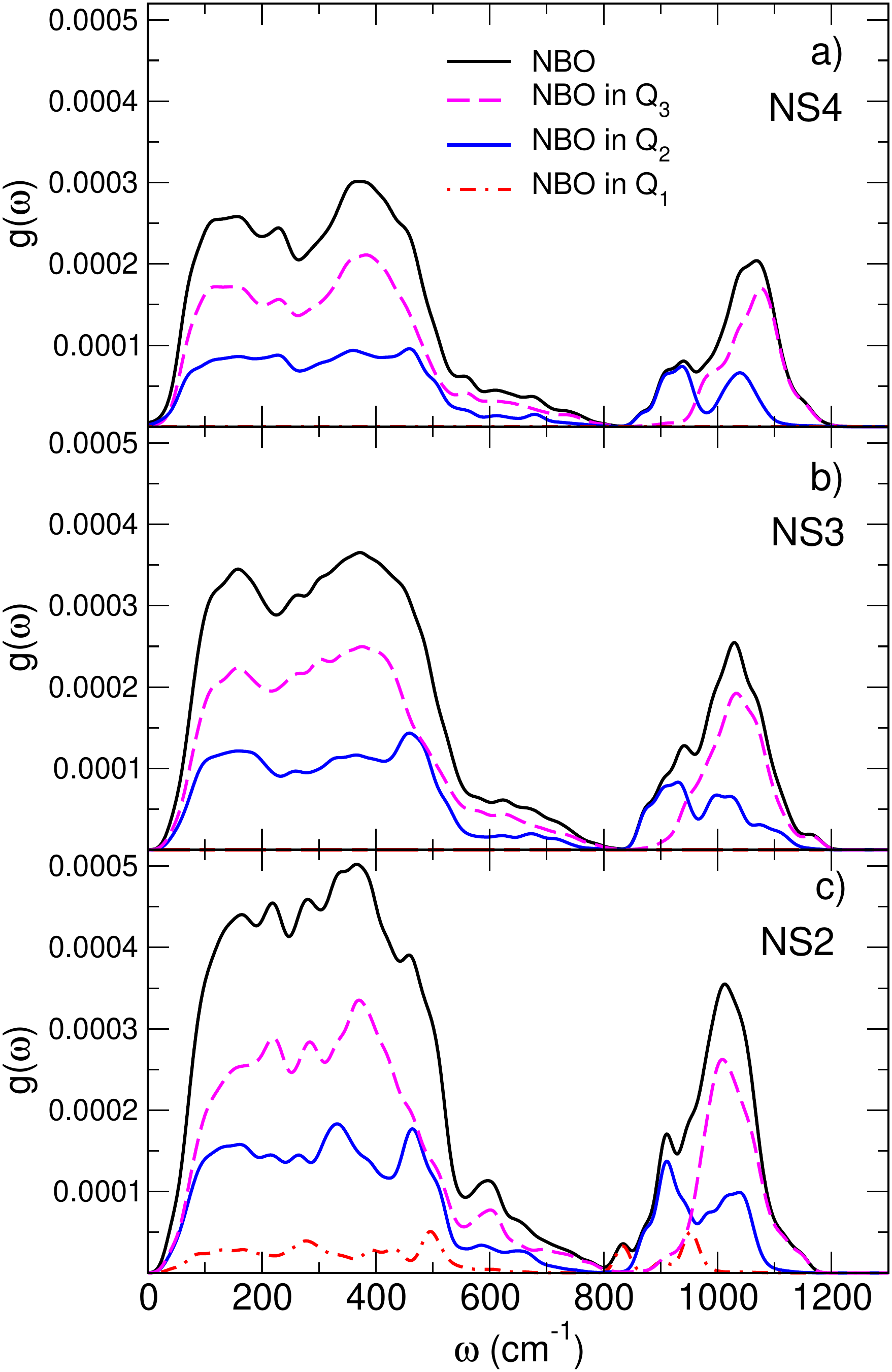} 
\caption{\label{fig:vdos_qnbo} Contributions to the NBO partial VDOS of  NBO's on Q$_2$ and Q$_3$-species for (a) NS4, (b) NS3, and (c) NS2.}
\end{figure}

A decomposition of the BO partial VDOS according the Si-BO-Si angle can link the vibrational properties with the geometry of the glass matrix. Such a decomposition is presented in Fig.~\ref{fig:vdos_ang}, where we see that the characteristics of the  BO contributions clearly change with respect to their Si-BO-Si  angle. 
The partial VDOS of BOs with angles between  140$^\circ$ and 160$^\circ$ dominates the partial VDOS of all BOs, and then, to a lower extent, comes the contribution of
BOs with angles between  120$^\circ$ and 140$^\circ$. The latter present a slightly higher intensity only for frequencies above 600~cm$^{-1}$ and below 850~cm$^{-1}$. 
Finally, we can identify the lowest contribution coming from BOs belonging to open Si-BO-Si angles ($160^\circ-180^\circ$). For these BOs, there is almost no mode in the frequency range between 600~cm$^{-1}$ and 850~cm$^{-1}$. 
The relative intensities of these three types of BOs  hint the relative concentrations of BOs with angles inside the given ranges, with the most probable Si-O-Si angles having values between  140 and 160$^\circ$. 
A notable effect of the Si-BO-Si angle is located around 450-500~cm$^{-1}$, where we observe an upshift of the smaller angles.
The inverse effect is present in the high-frequency band, where a decrease of the Si-O-Si angle produces a downshift of the average contribution to the VDOS. In this case, the downshift can be straightforwardly linked to stretching in  Q$_n$ species since, as will be discussed in Section~\ref{sec:nsx_struct}, smaller Si-O-Si angles are expected as the amount of NBOs on the Si atoms increases.

Finally,  we have investigated the contribution from the NBOs by  decomposing it with respect to the type of Q$_n$ species that they are linked to. Figure~\ref{fig:vdos_qnbo} shows the  contributions to the VDOS from NBOs on Q$_2$ and Q$_3$ entities. We remark that NBOs on Q$_2$ demonstrate a significantly higher intensity, compared to those on Q$_3$ for all three glasses. However, the most important difference between the two cases can be identified in the high-frequency band, where in all systems we observe a bimodal contribution for NBOs on Q$_2$ species and an overlapping peak for Q$_3$. In the former case, the peak centers were found to be at the same frequencies, 920$\pm10$~cm$^{-1}$ and 1020$\pm10$~cm$^{-1}$, whereas the maximum of the Q$_3$ contribution shifts from 1060 to 1010~cm$^{-1}$ when going from NS4 to NS2. We should also note that we do not comment on the contribution to the VDOS from NBOs on Q$_1$ entities, since only one exists in the NS2 system and therefore good statistics are not ensured.


\subsection{\label{sec:raman}Raman spectra}
The Raman spectra have been obtained by applying uniform electric fields and calculating 
the second order derivatives of the electron density matrix \cite{Lazzeri2003a}. 
The Raman susceptibility for each mode $p$ is given by:

\begin{equation} \label{eq:susc}
\mathcal{R}_{ij}^p = \sqrt{V} \sum_{I=1}^N \sum_{k=1}^3 \frac{\partial \chi_{ij}}{\partial R_{I,k}} 
\frac{{\mathbf e}_{I,k} (\omega_p)}{\sqrt{M_I}}
\end{equation}

\noindent where $V$ is the system volume, $i, j, k  = 1, 2, 3$,  $M_I$ the atom $I$ mass, while 
${\mathbf e}_{I,k} (\omega_p)$ is the 3-components vector defined in Eq.~(\ref{eq:pvdos}).
In Eq.~(\ref{eq:susc}), the Raman tensor $\displaystyle\frac{\partial \chi_{ij}}{\partial R_{I,k}}$ is given by:

\begin{equation}
\frac{\partial \chi_{ij}}{\partial R_{I,k}} = \frac{1}{V} 
\frac{\partial^2 F_{I,k}}{\partial \mathcal{E}_i \mathcal{E}_j} \bigg\rvert_{\mathcal{E}=0}
\end{equation}

\noindent with $F_{I \gamma}$ the force exerted on atom $I$ and  $\mathcal{E}$ the applied electric field. 
For the polarized Raman spectra (VV), the intensity for each mode $p$ is given by: 

\begin{equation}
I_{VV}^p = \alpha_p^2 + \frac{4 \tau_p^2}{45}
\label{eq:Ivv}
\end{equation}

\noindent where $\alpha_p$ and $\tau_p$ are the trace and anisotropy of the susceptibility:

\begin{equation}
\alpha_p = \frac{\mathcal{R}_{11}^p + \mathcal{R}_{22}^p + \mathcal{R}_{33}^p}{3},
\label{eq:susc_trace}
\end{equation}

\begin{widetext}
\begin{equation}
\tau_p^2 = \frac{\left(\mathcal{R}_{11}^p-\mathcal{R}_{22}^p\right)^2 +
 \left(\mathcal{R}_{22}^p-\mathcal{R}_{33}^p \right)^2 + \left(\mathcal{R}_{33}^p-
 \mathcal{R}_{11}^p \right)^2 + 6 \left[\left(\mathcal{R}_{12}^p\right)^2 + \left(\mathcal{R}_{13}^p\right)^2 + 
 \left(\mathcal{R}_{23}^p\right)^2 \right]}{2},
 \label{eq:susc_aniso}
\end{equation}
\end{widetext}

\noindent The depolarized Raman spectra (VH) have been calculated by multiplying the VV spectra by the  depolarization ratio $\rho_{dep}^p$: 

\begin{equation} \label{eq:depol}
I_{VH}^p = \rho_{dep}^p  \cdot I_{VV}^p \quad \mbox{ with } \quad \rho_{dep}^p = \frac{3 \tau_p^2}{45 \alpha_p^2 + 4 \tau_p^2}
\end{equation}

In order to decompose the Raman spectra into  partial contributions 
 of a specific group of atoms, we have  written the intensity of mode $p$ as a sum of three terms:

\begin{equation}
I_{VV}^p =\sum_{\alpha}I_{VV,\alpha}^p + I_{VV,comp}^p + I_{VV,over}^p
\end{equation}

\noindent where only the atoms of type $\alpha$ are considered in Eq.~(\ref{eq:susc}) prior to the calculation of the partial intensities  $I_{VV,\alpha}^p$ using Eqs.~(\ref{eq:Ivv})--(\ref{eq:susc_aniso}) (i.e., $\alpha=$~Si,~O,~Na or $\alpha=$~BO,~NBO). The second and third terms  are respectively the contributions 
from the remaining atoms in the system (i.e. the ones of Si and Na if  $\alpha=$~BO,~NBO) and the overlapping (or interference) terms, due to the squared 
terms in Eqs.~(\ref{eq:Ivv})-(\ref{eq:susc_aniso}). As already pointed out by Umari and Pasquarello for the case of pure SiO$_2$ \cite{umari2003first}, even if the sum of the partial Raman spectra does not recover the total spectrum, as in the case of the VDOS, it is reasonable to assume that these partial spectra  give a measure of the relative ones.

\begin{figure} [htbp]
\centering
	\includegraphics[width=15cm]{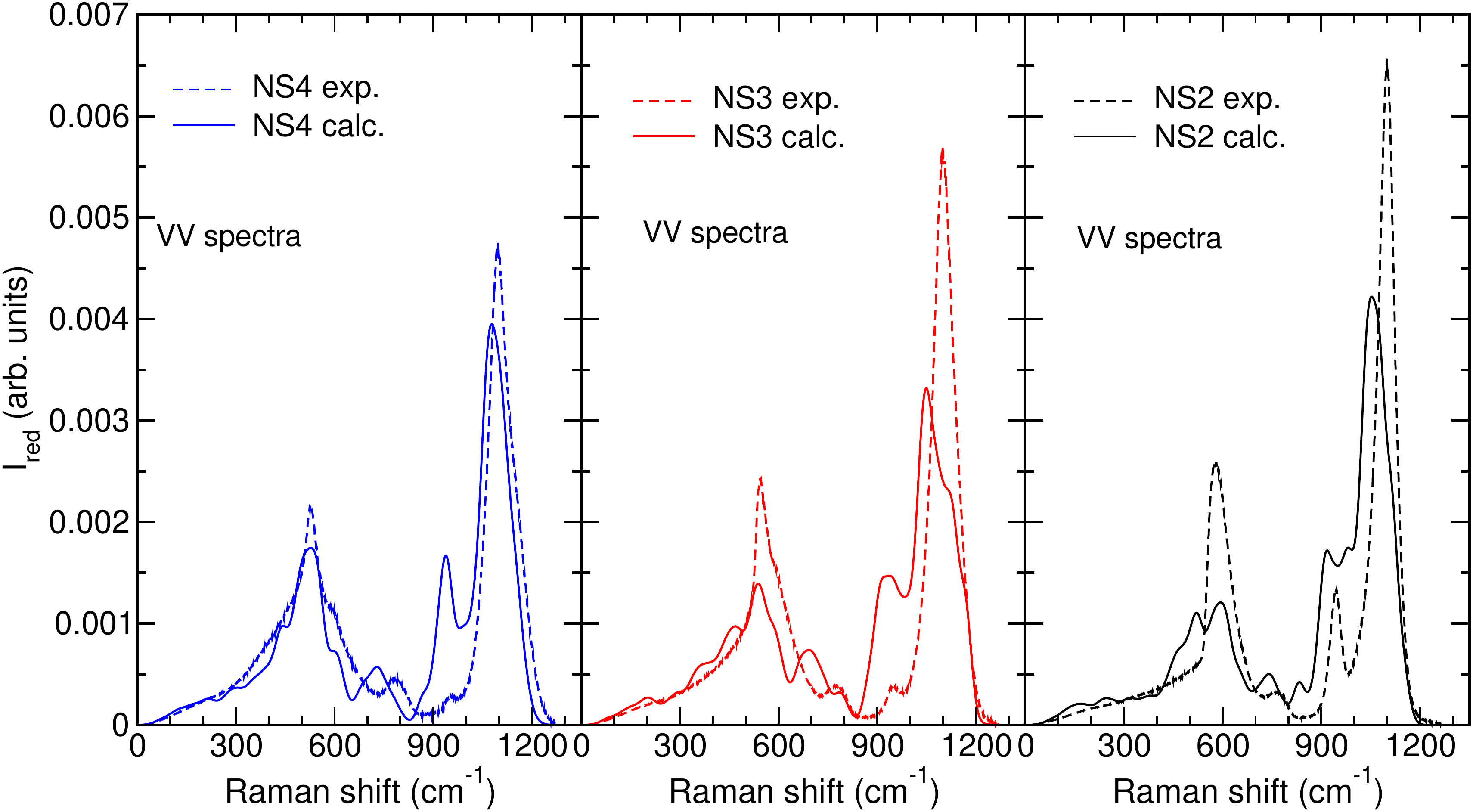}
	\includegraphics[width=15cm]{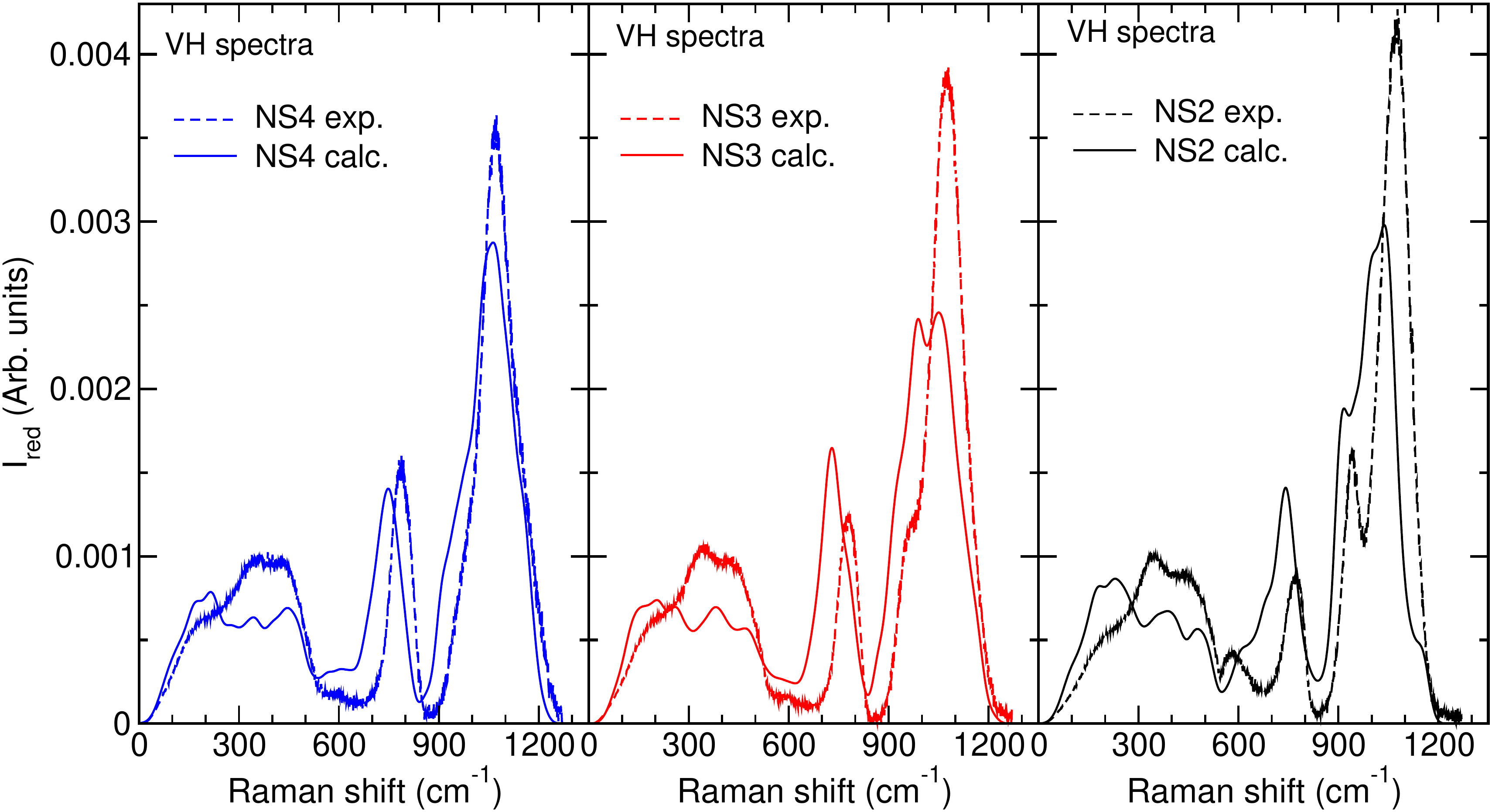}
\caption{\label{fig:raman_vvvh_all} Polarized (top) and depolarized (bottom)
 calculated Raman intensities for the three glasses (solid lines), 
alongside experimental results (dashed lines)~\cite{Hehlen2015}. 
The experimental  spectra are the reduced ones with respect to the measured intensities (see text for details).  The calculated as well as the experimental spectra are normalized to unity. }
\end{figure}

The VV and VH  calculated Raman spectra of the glasses under 
study are given in Fig.~\ref{fig:raman_vvvh_all} and compared to recent  experimental results~\cite{Hehlen2015}. 
The calculated spectra result  directly from the Gaussian broadening of the discrete intensities [see Eqs.~(\ref{eq:Ivv}) and (\ref{eq:depol})], while the experimental ones have been obtained by multiplying the experimentally measured intensities  by $\omega/\left[n\left(\omega\right)+1\right]$, where $n\left(\omega\right)=1/\left[\exp\left(\omega/k_BT\right)-1)\right]$ 
is the Bose factor, $k_B$ the Boltzmann constant, and $\mathrm{T=300~K}$. 
{In the Supplemental Material (see Fig.~S1)~\cite{suppl_mat}, 
 we also provide a comparison between the calculated spectra and the experimental ones, using an alternate definition for the reduced intensities. In that case, the latter have been multiplied by $1/[\omega (n(\omega)+1)]$, which is equivalent to dividing the calculated spectra in Fig.~\ref{fig:raman_vvvh_all} by $\omega^2$.

For either VV and VH spectra, the positions of the main bands are well reproduced by the calculated spectra  but there exist discrepancies in terms of band intensities.
If we consider the band between 500-600~cm$^{-1}$ in the VV spectra, its form and intensity is well reproduced for the highly polymerized (high silica content) NS4 composition.  
As it will be discussed later, this band contains the signal from Si-O-Si vibrations, i.e. the BO atoms. 
The calculated band is broader with respect to the experimental one and it is not narrowing  with increasing sodium content. This indicates that there are differences in the local and medium range structures present in our glass models and those in the real glasses. These differences are likely due  to the  small size of our glass models, which in the case of NS2 contains considerably less BOs and therefore poorer statistics compared to NS4. 
Further, the  discrepancies  for the high-frequency band (900--1200~cm$^{-1}$), corresponding to the response from the Q$_n$ species are due to the fact that observed Q$_n$ distributions are not exactly reproduced in our models. 
For the VH spectra, we see a broad band below 
600~cm$^{-1}$ (see Fig.~S1 
in Supplemental Material~\cite{suppl_mat}), which was assigned to two types of  alkali motion~\cite{Hehlen2015}, one at $\approx$175~cm$^{-1}$ and the second one at   $\approx$350~cm$^{-1}$. In spite of the small size of our models, we can identify in the calculated VH spectra the presence of small peak at  $\approx$350~cm$^{-1}$ showing an increasing intensity with increasing sodium content.

\begin{figure} [htbp]
	\centering
	\includegraphics[width=7cm]{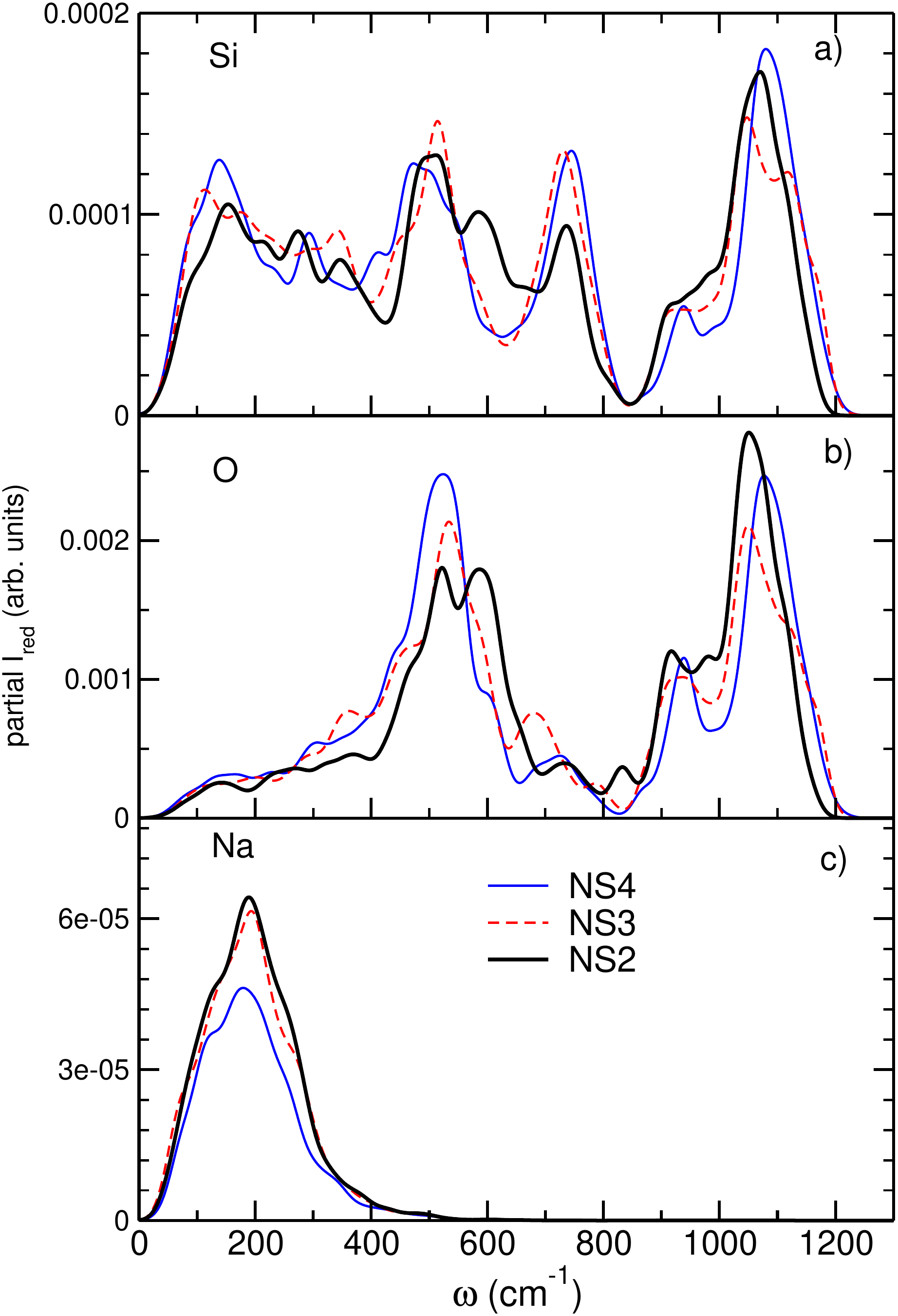}
	\caption{\label{fig:raman_decomp_species} Partial contributions of three constituent species: silicons (a), oxygens (b), and sodiums (c).}
\end{figure}

As for the pure VDOS, we have initially decomposed the VV spectra with respect to the contributions of the three species, Si, O, and Na and present the results in Fig.~\ref{fig:raman_decomp_species}.
The partial contributions stemming from the O atoms are by far the most significant, as it was already noticed for pure silica \cite{umari2003first}, as well as in the partial VDOS of our three sodosilicates (see Fig.~\ref{fig:fig2-vdos_all}).
This trend  is far more substantial in the polarized Raman spectra.
Consequently, there is strong similarity between the form of the oxygen contributions plotted in Fig.~\ref{fig:raman_decomp_species}(b) and the total VV Raman spectra shown in Fig.~\ref{fig:raman_vvvh_all}. With increasing sodium content, it is the oxygen contribution exhibiting the most noticeable changes, namely, a shift to the right of the band between 400--700~cm$^{-1}$ and a small shift to the left of the high-frequency band. 

	\begin{figure*}[htbp]
		\centering
		\includegraphics[width=12cm]{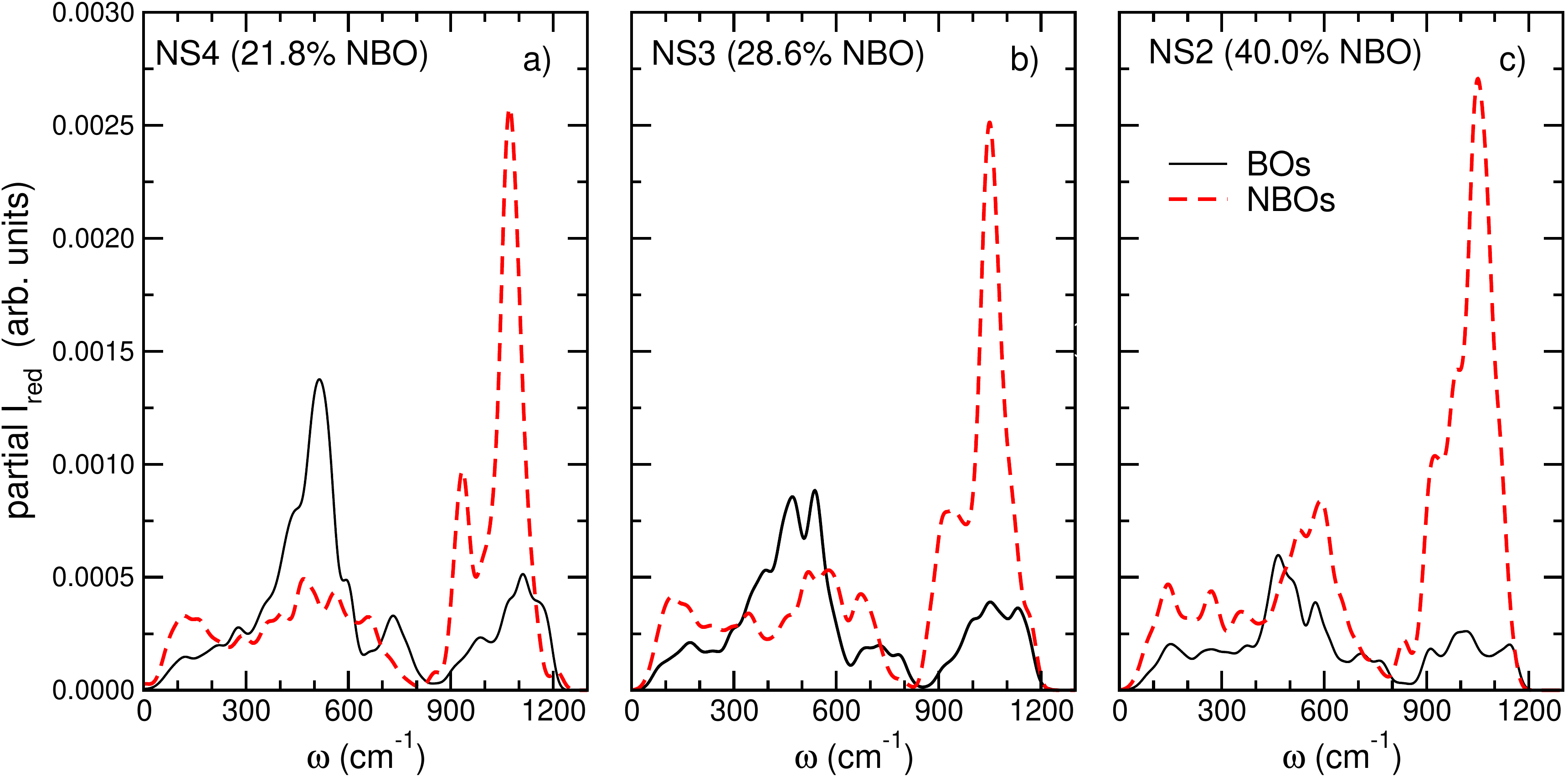}
		\caption{\label{fig:raman_bo-nbo}  Partial contributions to the  VV spectra of BO (solid lines) and NBO (dashed lines) atoms for a) NS4, b) NS3 and c) NS2.}
	\end{figure*}

Due to the prominent contribution of oxygen vibrations to the Raman spectra, we have further considered three  decompositions: according to their species (BO or NBO), the bond angle of the Si-O-Si bridgings,  and the contribution of the NBOs  to the high frequency band.
The decomposition according to the contributions from BOs and NBOs is given in Fig.~\ref{fig:raman_bo-nbo}. 
We initially point out the strong contribution of the NBOs to the high-frequency, or Q$_n$,  band. This happens in spite of their smaller concentration with respect to the BOs in any of the three compositions, suggesting that they are the most Raman active in this band.
The intensity of these modes largely dominates the one of the oxygens in Si-O-Si bridges, that one finds in the pure silica glass.   The vibrations of NBOs  then  define the shape and intensity of this Raman band, and we  expect that they  are predominantly of
stretching type~\cite{zotov1999calculation}. 
For the rest of the spectra (below 800~cm$^{-1}$) the NBOs present a distinct signal whose intensity increases with increasing sodium content, and surpasses the one of BOs in the case of NS2.
	

As for BOs, they present a significant band between 400--600~cm$^{-1}$, particularly for low sodium content.
Taking into account the decomposition of the pure VDOS with respect to rocking, bending and stretching motions of the Si-O-Si bridges (see Fig.~\ref{fig:vdos_rbs}), the Raman BO vibrations could be assigned either to rocking or bending motions, or both. However, for pure SiO$_2$, Umari and Pasquarello \cite{umari2003first} have shown that the VV spectrum below 900~cm$^{-1}$ is dominated by bending motions, while the rocking contribution is almost suppressed, in spite of their large weight in the pure VDOS. This conclusion likely holds also for sodosilicate compositions. It is supported by the fact that  the relative contributions of the rocking, bending, and stretching vibrations [see Figs.~\ref{fig:vdos_rbs}(d)--\ref{fig:vdos_rbs}(f)]   do not seem to show a strong dependence on sodium content, on one side, while their frequency dependence also presents a  close resemblance to the one reported for $a$-SiO$_2$ (see Ref.~\cite{umari2003first}). Finally, if we consider the   contribution of the BOs  to the Q$_n$ band, our calculations show that the presence of sodium leads to the suppression of the double-peak feature, well known for pure SiO$_2$ glass.

 \begin{figure} [htbp]
 	\centering
 	\includegraphics[width=7cm]{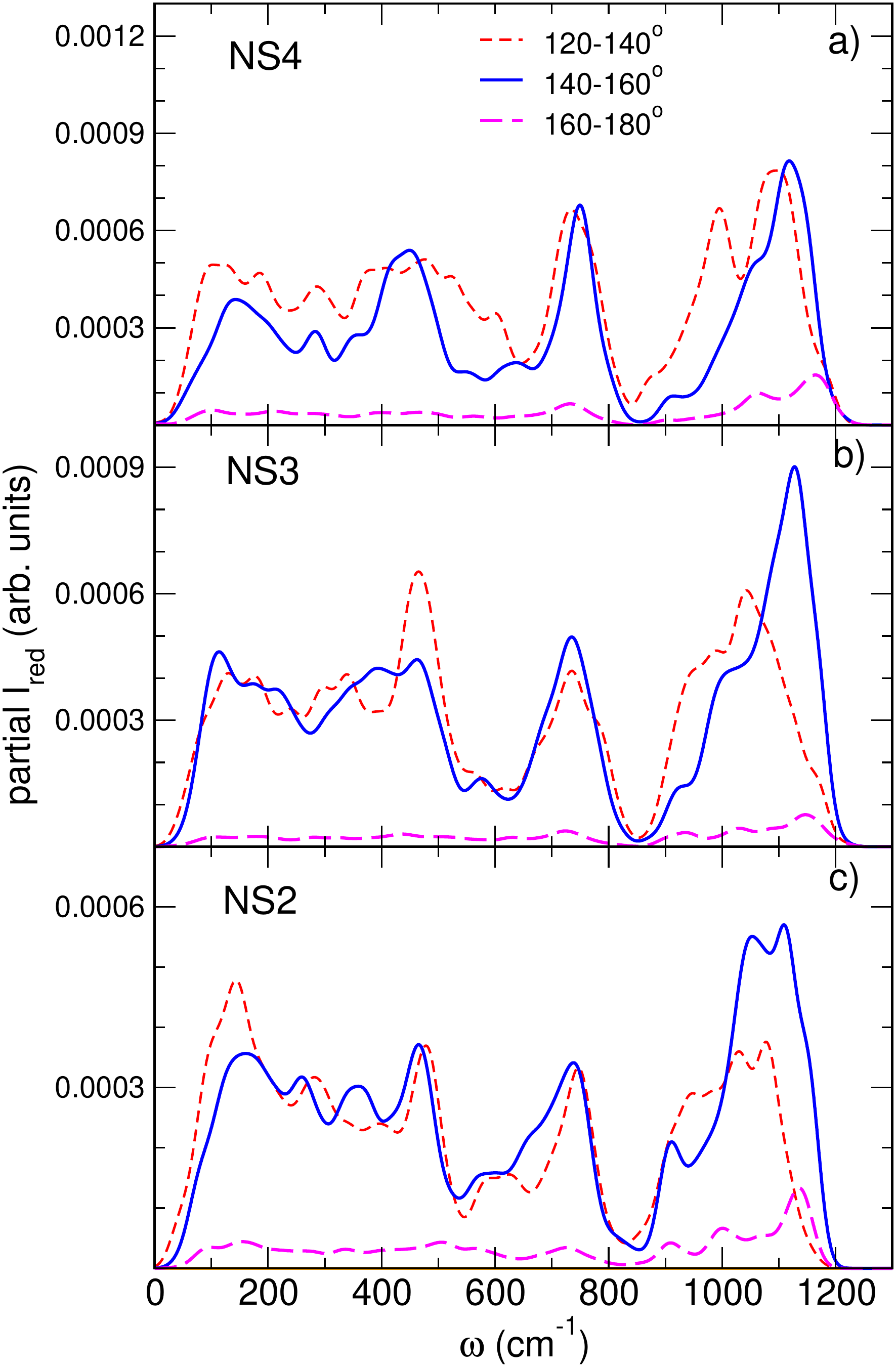}
 	\caption{\label{fig:raman_angle} 
 		Contributions to the polarized Raman spectra of the BO atoms taking into account the value of the Si-O-Si bond angle   for a) NS4, b) NS3 and c) NS2. }
 \end{figure}
 
The analysis of the polarized Raman spectra according to the Si-O-Si angle can be considered, to a certain extent, as a further decomposition of the bridging oxygen contribution with respect to its local environment.  As for the pure VDOS, we have considered three angular ranges, (120$\degree$-140$\degree$, 140$\degree$-160$\degree$, and 160$\degree$-180$\degree$, respectively) and the resulting partial contributions are given in Fig.~\ref{fig:raman_angle}. 
In the low and mid-frequency bands (below 850~cm$^{-1}$) we observe that the partial contribution of the small  angles, 120$\degree$-140$\degree$, has almost the same weight as the one of larger angles, between 140$\degree$ and 160$\degree$,  and this despite their percentage is  smaller that the one of larger angles. 
These quite similar contributions to the Raman spectra may arise from the variation of the Raman coupling factor which linearly decreases with increasing Si-O-Si angle (see Fig.~S2 
 in the Supplemental Material~\cite{suppl_mat} as well as Refs.~\cite{Giacomazzi2009,umari2003concentration}).
 For the high frequency band,  these two main contributions  show a stronger overlap with respect to the ones found in the pure VDOS.  The contribution of BOs with angles between  140$\degree$-160$\degree$ is rather asymmetric and seems to become stronger with increasing Na content for the highest frequencies, with respect to the one of BOs with angles smaller than 140$\degree$. Finally we note that the contribution of the very large angles, above 160$\degree$, is almost negligible below 850~cm$^{-1}$, and seems to slightly increase with increasing frequency in the Q$_n$-band. 

\begin{figure} [htbp]
	\centering
	\includegraphics[width=12cm]{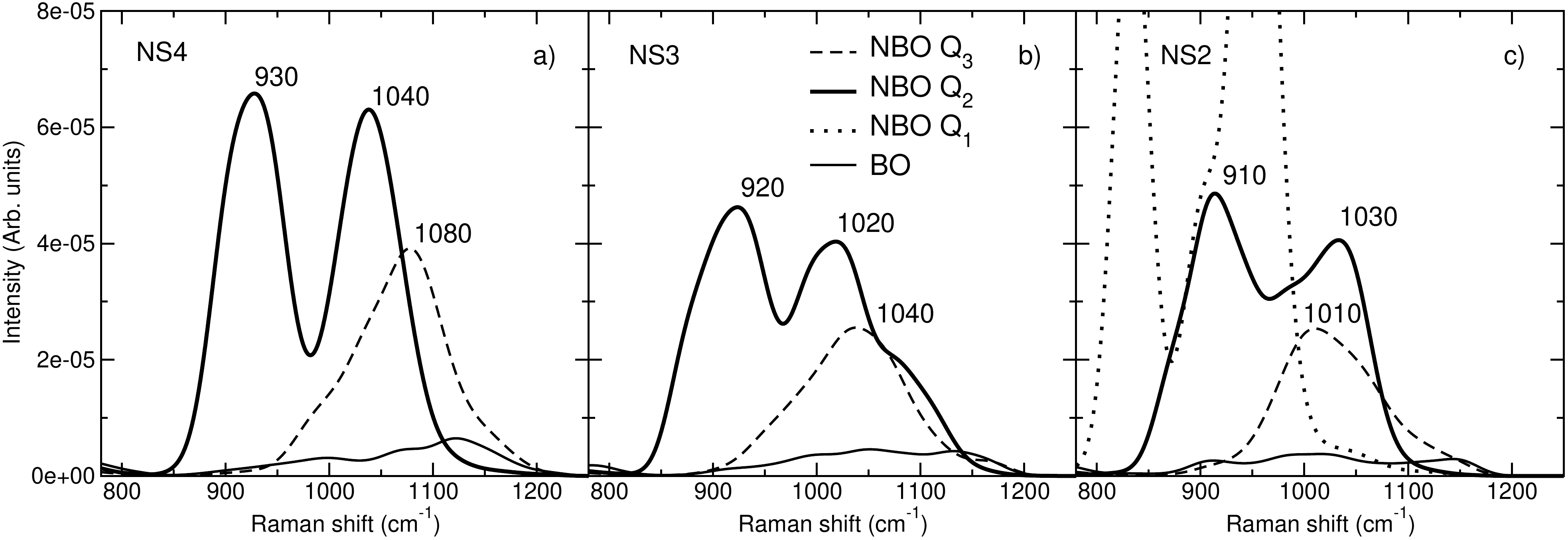}
	\caption{\label{fig:raman_qnbo} Per-atom contribution to the Raman spectra of NBO's on different Q$_n$-species for (a) NS4, (b) NS3, and (c) NS2. The per-atom contribution of BO's is also given for comparison. The maxima of the contributions for Q$_2$ and Q$_3$ species are noted on the plots.}
\end{figure}

An investigation of the per-atom NBO contributions in the high-frequency band, given in Fig.~\ref{fig:raman_qnbo}, shows that the Q$_n$ contributions are significantly overlapping.
 Regarding the NBOs on Q$_1$ species, the poor statistics caused by the single case observed in the NS2 sample makes us refrain from further comments. In the case of NBOs on Q$_2$ species, we observe a well-defined doublet structure with maxima around 920 and 1030~cm$^{-1}$, whose position does not seem to be strongly affected by composition. The most probable reason for the appearance of the doublet is the presence of neighboring effects from different Q$_n$ species linked to Q$_2$ ones. However, our analysis with respect to the populations of Q$_n$ neighbors did not reveal a clear correlation. As for the NBOs on Q$_3$ species, they exhibit a unimodal distribution whose maximum shifts from 1080~cm$^{-1}$ in NS4 to 1030~cm$^{-1}$ in NS2. The particular form of the NBO contributions for Q$_2$ and Q$_3$ species has been first pointed out by Zotov \textit{et al.} for NS4 \cite{zotov1999calculation} and is found to be in remarkable agreement to the more recent experimental works by Malfait \textit{et al.} for sodosilicates~\cite{Malfait2008} and Woelffel \textit{et al.} for soda-lime silicates \cite{Woelffel2015}. The Q$_4$ species give rise to a very broad band ranging from 900 to 1200~cm$^{-1}$, underneath the Q$_2$, Q$_3$ contributions. 
The strong overlaps between the different Q$_n$ entities, as well as the form of the individual contributions, indicate that the quite common decomposition of experimental data with gaussian peaks should be considered cautiously.

\subsection{\label{sec:ns2_quench}Comparison of the NS2 samples}
It could be argued that the bimodal distribution of the Q$_2$ contribution to the Raman spectra results from the relatively low total number of these units which lead to poor statistics. In order to clarify this point, we present in Fig.~\ref{fig:ns2_all} the total polarized Raman spectra for the two NS2 systems prepared with a quench rate of 10$^{11}$~K/s (henceforth named model I and model II), alongside the one with 10$^{10}$~K/s which has already been discussed. Models I and II have very similar percentages of Q$_2$ and Q$_3$ species, 26.1\% Q$_2$ and 28.3\% Q$_3$ for model I, 24.4\% Q$_2$ and 31.1\% Q$_3$ for model II, which reflects to an almost identical structure of the Q$_n$ band. On the contrary, the NS2 glass prepared with the 10$^{10}$~K/s quench rate has a considerably lower amount of Q$_2$ entities, hence the lower signal around 950~cm$^{-1}$. Furthermore, the partial contributions from NBOs on Q$_2$ (inset of Fig.~\ref{fig:ns2_all}) exhibit a clearly bimodal distribution for all three systems, suggesting that the statistics are satisfactory. Small differences in the form of the distribution and the positions of the maxima are expected to derive from slightly different local environments around the Q$_2$ units. The comparison between the three NS2 models also corroborates the relatively high cross-section from the Q$_2$ species, since a decrease of approximately 8\% in their population leads to an almost 50\% intensity decrease of the 950~cm$^{-1}$ peak. 

Additionally, the similar forms and positions of the different contributions for all three models indicate that the accurate prediction of the form of the Q$_n$ band is principally a question of obtaining the correct distributions of the Q$_n$ species in the computational models and that the approximations within the DFT calculations are of lesser overall importance. Further refinement of the glass preparation protocols would open the way for a systematic modeling of the Raman spectra for silicate glasses and a rigorous interpretation of experimental data.

\begin{figure} [htbp]
\centering
\includegraphics[width=8cm]{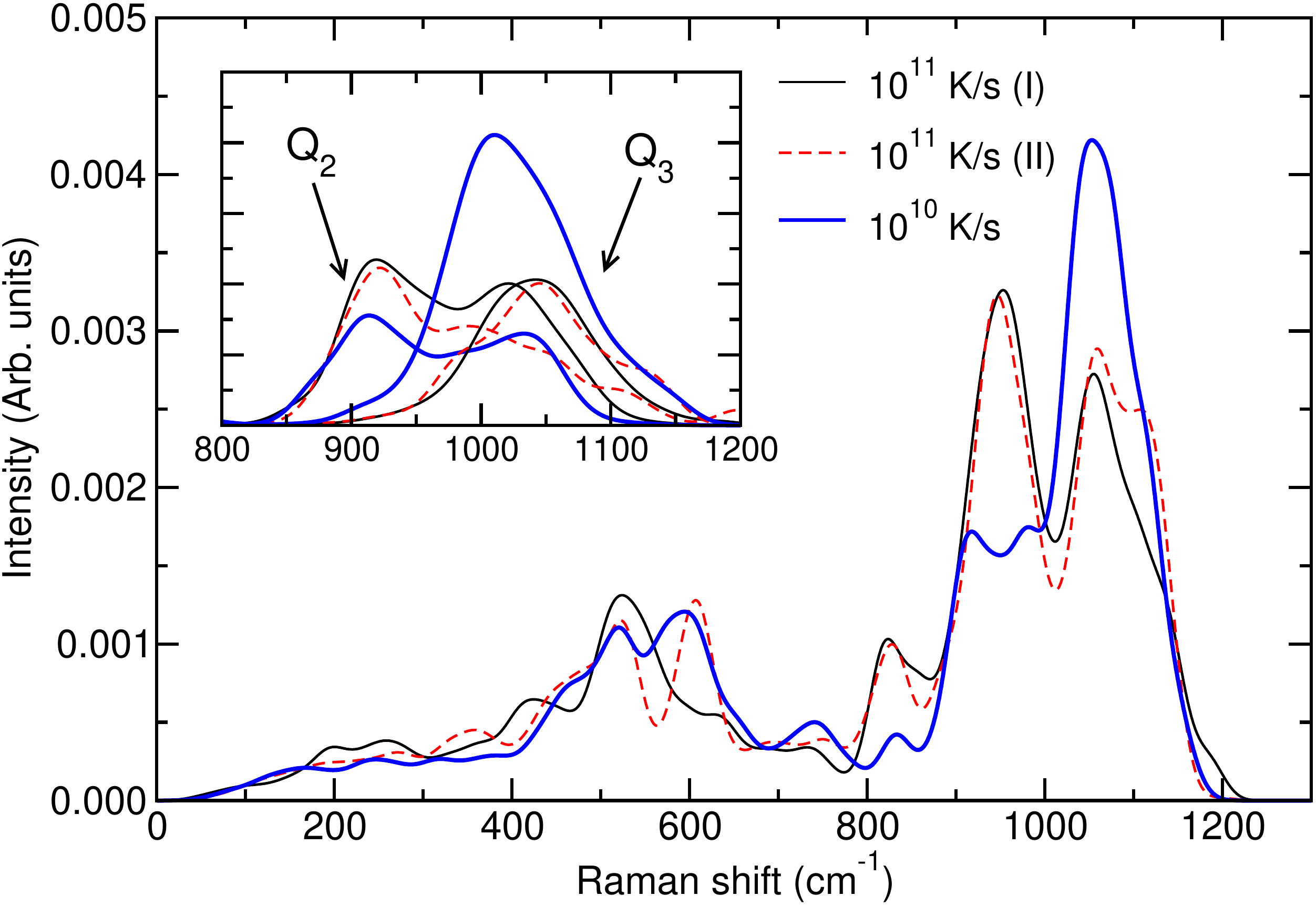}
\caption{\label{fig:ns2_all} Polarized Raman spectra of the three NS2 samples obtained with different quench rates. Inset: total contribution from NBO's on Q$_2$ and Q$_3$ species for the three models.}
\end{figure}

\section{Conclusion}
We have presented in this work the investigation of the vibrational spectra, VDOS and Raman, for three sodosilicate glasses. The visualization of the vibrational modes has been used to make an initial analysis in order to show that the experimentally predicted modes are reproduced in the calculations. 
The per-species decomposition of the VDOS revealed the relatively high contribution from O atoms in the whole frequency range, and the contribution of the Na atoms at low frequencies, typically below 400~cm$^{-1}$. 
The mode analysis of the Si-O-Si bridges has shown that the vibrations up to 800~cm$^{-1}$ are of rocking and bending type, whereas above that frequency they are essentially stretchings. The average contributions of the bridgings to the VDOS seems to be affected by the angle formed by the three atoms, with a slight upshift around 500~cm$^{-1}$ and a downshift at the high-frequency band as the angle becomes smaller. Concerning the high-frequency band, the results reveal important overlaps between NBO contributions in Q$_2$ and Q$_3$ species, as well as with those of the BOs.  

The Raman spectra reproduce well the experimental band positions, whereas the calculated intensities show some deviations. We have shown that this is due to the limited system size and the prediction of the Q$_n$ species distribution by the classical potential which was used to quench the glasses from the melt. The calculation of the Raman response by Si-O-Si bridges has shown that large angles have a particularly flat contribution to the lower and middle parts of the spectra while, similarly to the VDOS, relatively small angles exhibit an upshift around 500~cm$^{-1}$. The calculation of the BO and NBO partials to the Raman spectra clearly shows that the form of the Q$_n$ band is defined by the latter. The further analysis of the NBOs in terms of their local environment shows a bimodal contribution with well-defined maxima for those connected to Q$_2$ and a highly overlapping Q$_3$ peak. The fact that the BO partial is also overlapping and spans the entire width of the Q$_n$ band indicates that one has to be particularly prudent when deconvoluting experimental spectra with the use of simple functions.

\section{Acknowledgements}
D.K. would like to thank CEA for the financial support.
This work was granted access to the HPC resources of TGCC/CINES/IDRIS under the allocation x2016097572 attributed by GENCI (Grand Equipement National de Calcul Intensif). 


%


\clearpage


\beginsupplement

\section*{\label{sec:supplmat}Supplementary material}

\centerline{\textbf{D. Kilymis, S. Ispas, B. Hehlen, S. Peuget and J.-M. Delaye}}

\bigskip
\noindent
List of some videos corresponding to atomic vibrations at selected frequencies:  

\begin{itemize}
	
	\item  Video 1: Motions of rigid tetrahedra and dangling-like motions of Na-O bonds ($\omega=108$~cm$^{-1}$).
	\item  Video 2: Breathing motions of oxygens atoms (highlighted in blue) in a 4-member ring, as well as bending motions of Si-O-Si bridges ($\omega=477$~cm$^{-1}$).
	\item  Video 3: Breathing motions of oxygens atoms (highlighted in blue) in a 3-member ring ($\omega=610$~cm$^{-1}$). 
	 	\item  Video 4: Streching-like motions of oxygen atoms in high-frequency band ($\omega=1130$~cm$^{-1}$) 
\end{itemize} 
These  videos were produced using Avogadro version 1.1.1~\cite{avogadro1,avogadro2}, and are available upon request from the authors.

\begin{figure} [h]
\centering
\includegraphics[width=14cm]{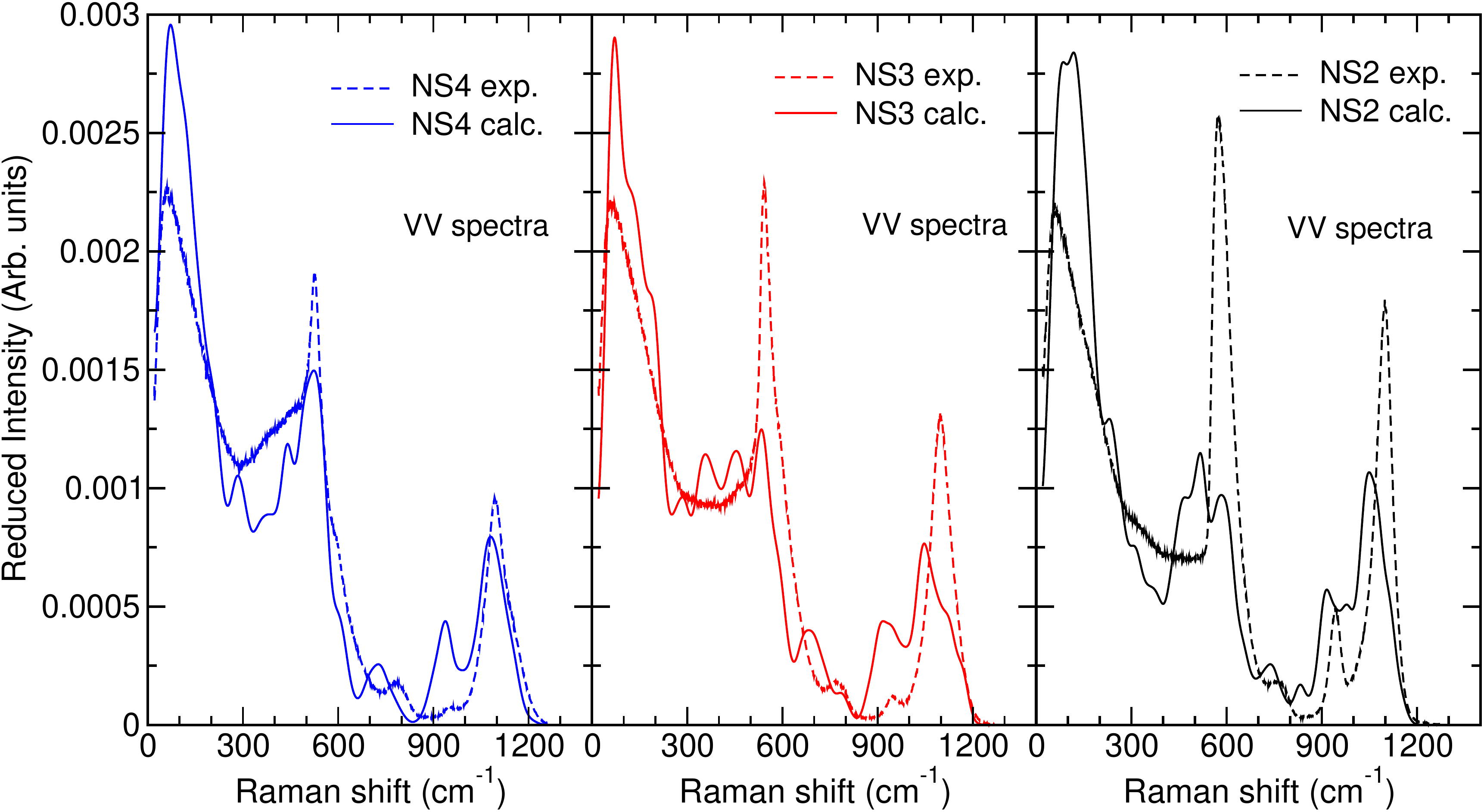}

\includegraphics[width=14cm]{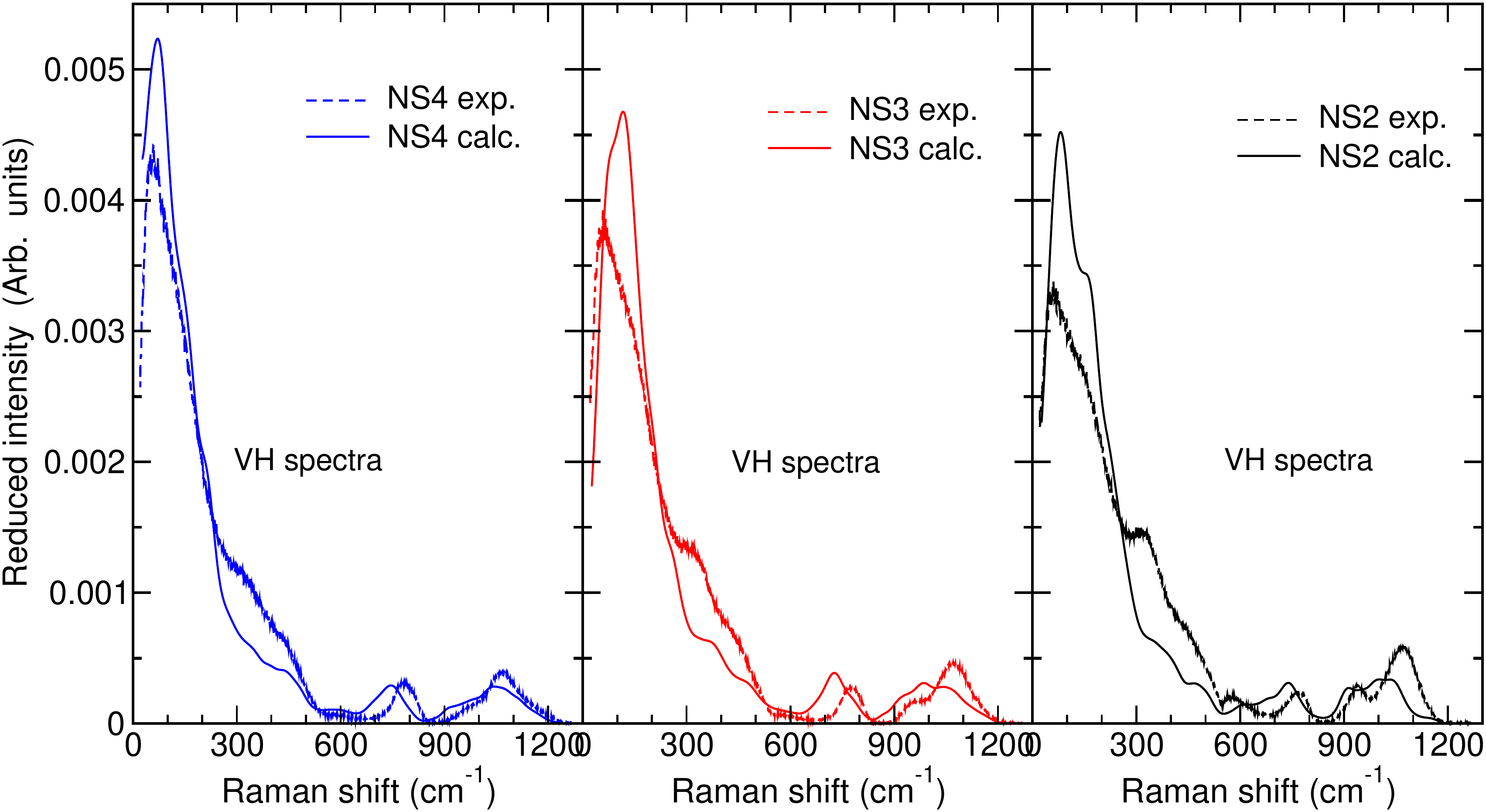}

\caption{\label{fig:raman_vvvh_all-normBH} a) Reduced VV (top panel) and VH  (bottom panel) Raman spectra for the three glasses (solid lines), alongside experimental results (dashed lines) \cite{Hehlen2015}. The experimental reduced spectra were  multiplied by $1/[\omega\left(n\left(\omega\right)+1\right)]$, where $n\left(\omega\right)=1/\left(\exp\left(\omega/k_BT\right)-1)\right)$ 
is the Bose factor, $k_B$ the Boltzmann constant and $\mathrm{T=300~K}$. The calculated as well as the experimental spectra are normalized to 1.}
\end{figure}

\begin{figure} [htbp]
	\centering
	\includegraphics[width=12cm]{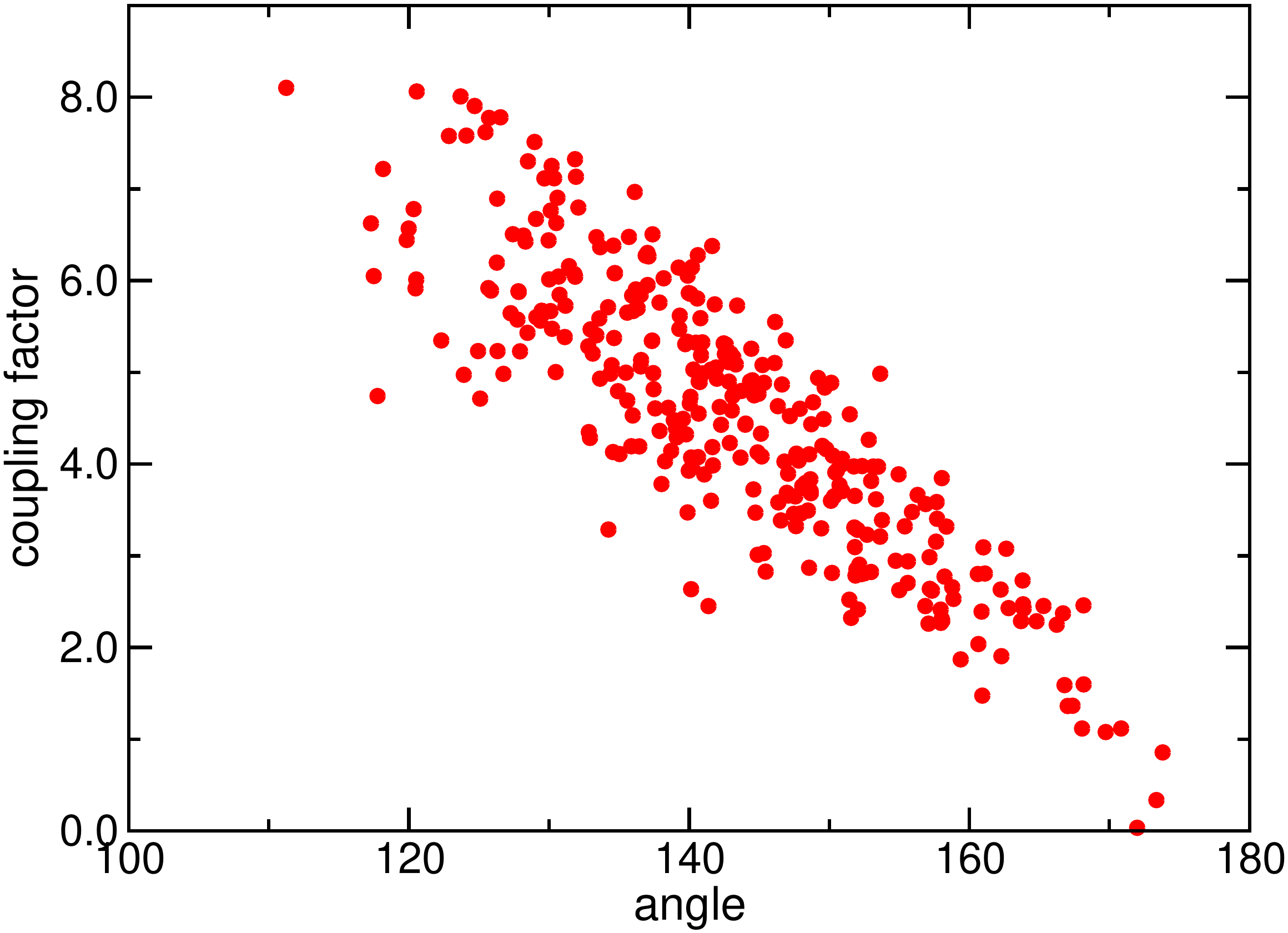}
	\caption{\label{fig:coupling_factor} Angular dependence of the coupling factor \cite{Giacomazzi2009,umari2003first} calculated for the models studied in this work. As no compositional dependence has been detected, we have made no distinction between the different samples.
}
\end{figure}

\end{document}